\documentstyle[psfig]{mn}

\newif\ifAMStwofonts

\def\etal{{\it et al. }}

\begin{document}

\title[Hickson Compact Group Galaxies]{Ages and Metallicities of Hickson Compact Group Galaxies}
\author[R. Proctor et al.]
{Robert. N. Proctor$^{1}$, Duncan. A. Forbes$^{1}$, George. K. T. Hau$^{2}$, Michael. A. Beasley$^{1}$,\\
\\
\LARGE G. M. De Silva$^{1}$, {R. Contreras$^{3}$ and A. I. Terlevich$^{4}$}\\ 
$^1$ Centre for Astrophysics \& Supercomputing, Swinburne University,
Hawthorn VIC 3122, Australia\\
Email: rproctor@astro.swin.edu.au, dforbes@swin.edu.au\\
$^2$ European Southern Observatory, Garching, Germany\\
$^3$ Departmento de Astronomia y Astrofisica, P. Universidad
Catolica, Casilla 104, Santiago 22, Chile\\
$^4$ School of Physics and Astronomy, University of Birmingham,
Birmingham,B15 2TT, UK\\ 
}

\pagerange{\pageref{firstpage}--\pageref{lastpage}}
\def\LaTeX{L\kern-.36em\raise.3ex\hbox{a}\kern-.15em
    T\kern-.1667em\lower.7ex\hbox{E}\kern-.125emX}

\newtheorem{theorem}{Theorem}[section]

\label{firstpage}

\maketitle

\begin{abstract}
Hickson Compact Groups (HCGs) constitute an interesting extreme in the range of
environments in which galaxies are located, as the space density of galaxies
in these small groups are otherwise only found in the centres of much larger
clusters. The work presented here uses Lick indices to make a comparison
of ages and chemical compositions of galaxies in HCGs with those in other
environments (clusters, loose groups and the field). The metallicity and
relative abundance of `$\alpha$-elements' show strong correlations with
galaxy age and central velocity dispersion, with similar trends found
in all environments. However, we show that the previously reported correlation between
$\alpha$-element abundance ratios and velocity dispersion disappears
when a full account is taken of the the abundance ratio pattern in the
calibration stars. This correlation is thus found to be an
artifact of incomplete calibration to the Lick system.

Variations are seen in the ranges and 
average values of age, metallicity and $\alpha$-element abundance ratios for
galaxies in different environments. Age distributions support the
hierarchical formation prediction that field galaxies are on average younger
than their cluster counterparts. However, the ages of HCG galaxies are 
shown to be more similar to those of cluster galaxies than those in the
field, contrary to the expectations of current hierarchical models. 
A trend for lower velocity dispersion galaxies to be younger was also seen.
This is again inconsistent with hierarchical collapse models, but is qualitatively 
consistent with the latest N-body-SPH models based on monolithic collapse in which star
formation continues for many Gyr in low mass halos.

\end{abstract}


\section{Introduction}

Hickson Compact Groups (HCGs) were defined by Hickson (1982) to
be spatially compact collections of four or more galaxies that
are relatively isolated from their general surroundings. The
original selection was made without knowing individual redshifts
of the galaxies. Consequently a few galaxies have been shown
to be merely chance projections. Although claims were made that 
such projections were a major contributor to the appearance of 
HCGs (e.g. Mamon 1986; Hernquist, Katz \& Weinberg 1995), the reality
of most HCGs is now accepted (e.g. Ponman \etal 1996; Tovmassian
\etal 2000). 

The space density of galaxies in HCGs is very high, and
comparable to that at the centres of
galaxy clusters. Furthermore the velocity dispersions within HCGs
are quite low ($\sim$200 km s$^{-1}$). These two conditions imply that
HCGs are the galactic environments in which the effects of
mergers and interactions should be most pronounced.
Signs of mergers and interactions have indeed been found in HCGs
(e.g. Rubin, Hunter \& Ford 1991; Forbes 1992; Mendes de Olivera
\& Hickson 1994; Jones, Ponman \& Forbes 2000). However, the merging 
rates and starburst/AGN activity seem to be lower than expected (Zepf 
\& Whitmore 1991; Coziol \etal 1997; Verdes-Montenegro \etal 1998). It
has also been shown (Ponman \etal 1996) that most HCGs are nearly or
fully virialized systems, indicating that most are fairly evolved. This
is surprising, given the short crossing times in most HCGs, as such groups
should be unstable and merge into a single massive galaxy within the Hubble
time (White 1990). These considerations have prompted speculation about the
formation and evolution of HCGs. One possibility is that the galaxies do 
merge, and form a single massive elliptical galaxy, but, with secondary 
infall of surrounding galaxies, the systems maintain  their status as HCGs 
(Governato, Tozzi \& Cavaliere 1996). A second possibility is that loose groups may be 
continously collapsing to form more compact groups (Diaferio, Gell \& Ramella 1994). 
Alternatively, simulations suggest that if the dark matter is distributed 
in a common group halo, rather than individual galaxy halos, merging is 
suppressed and HCGs would be dynamically long-lived systems (Athanassoula, 
Makino \& Bosma 1997).

It is also useful to consider the  predictions of the various models of 
galaxy formation. For instance, Kauffmann (1996) used a semi-analytical 
approach to modeling the hierarchical formation of galaxies in different 
environments, finding a bimodal age distribution in dark matter halos of 
size 10$^{13}$ M$_{\odot}$, which includes HCGs, small group and field 
galaxies. In such halos the  Kauffmann models predict the central 
luminous ellipticals to have younger average ages than their less luminous 
neighbours. On the other hand, the N-body-SPH monolithic 
collapse models of elliptical galaxy formation of Chiosi \& Carraro (2002) predict 
younger ages for low mass galaxies than for those with high mass. 

One way to discriminate between the different evolutionary models would be 
to determine ages and chemical compositions of a number of early-type galaxies 
in HCGs for comparison to those of galaxies in other environments. For example, 
are they  generally old (indicating a low interaction rate) like galaxies in 
the centre of clusters, or is there evidence of an enhanced interaction rate 
as indicated by the younger stellar populations often seen in loose group
and field galaxies?

In this paper we use an approach which has been
successfully applied to galaxies in clusters, loose groups and the
field, i.e. using Lick indices to obtain relative ages and
metalicities (e.g. Gonzalez 1993; Kuntschner 2000; Trager \etal 2000, hereafter
T00; Poggianti \etal 2001; Kuntschner \etal 2002; Proctor \& Sansom 2002,
hereafter PS02; Terlevich \& Forbes 2002). Here we extend these techniques
to HCGs for the first time. The results for HCGs are compared to galaxies
in different  environments and the predictions of galaxy formation models. 
We assume H$_0$=75 km s$^{-1}$ Mpc$^{-1}$ throughout.

\section{The Sample}

Groups were chosen from the Hickson Compact Group Catalog (Hickson 1982). 
The groups were HCG 4, 14, 16, 22, 25, 32, 40, 42, 62 and 86. 
The spectra of several galaxies not in HCGs were
also obtained. These include three field Arp-Madore galaxies NGC~2502, 
NGC~3203 and  NGC~6684, 2 galaxies in the massive loose group NGC~5044, and 
NGC~3305 located in the loose group LGG211 (Garcia 1993). 
Our emphasis was to obtain spectra for the early-type galaxies.  
However, a number of spiral galaxies were also observed. A number of
galaxies failed to achieve a suitable signal-to-noise and have been
omitted (see Section \ref{ind_meas}). The remaining galaxies exhibit
signal-to-noise ratios of between 20 and 60. These include a total of 17 
early-type galaxies and 9 spiral bulges in HCGs, and 6 early-type galaxies 
in groups.

\section{Observations and Initial Data Reduction}

\begin{table} 
\footnotesize 
\begin{center} 
\begin{tabular}{lccc} 
\hline 
 Galaxy      & Date    & Exp. time & P.A.      \\ 
             &         & (secs)    & (degrees) \\
\hline
{\bf 2000 Apr}&{\bf Long-slit}             &       &     \\
HCG62:ZM19   & 2000/04/09  & 3x600 & 117 \\
NGC~5046      & 2000/04/11  & 2x600 & 50  \\
NGC~5049      & 2000/04/11  & 4x600 & 33  \\
NGC~6684      & 2000/04/09  & 3x600 & 125 \\
\hline
{\bf 2000 Apr}&{\bf MOS}            &        &     \\
HCG42A       & 2000/04/09 & 3x1200 & 108 \\
HCG42C       & 2000/04/09 & 3x1200 & 108 \\
HCG86A       & 2000/04/11 & 3x1200 & 118 \\
HCG86B       & 2000/04/11 & 3x1200 & 118 \\
\hline
{\bf 2001 Jan}&{\bf Long-slit}&       &     \\
HCG62:ZM22   & 2000/12/30 & 3x600 & 80  \\
HCG40B       & 2000/12/31 & 3x600 & 10  \\
NGC~2502      & 2001/01/01 & 3x600 & 30  \\
NGC~3203      & 2001/01/01 & 2x600 & 330 \\
NGC~3305      & 2001/01/01 & 3x600 & 270 \\
\hline
{\bf 2001 Jan}&{\bf MOS}&        &   \\
HCG4A     & 2000/12/30 & 3x1200 & 0 \\
HCG4C     & 2000/12/30 & 3x1200 & 0 \\
HCG14A    & 2001/01/01   & 3x1200 & 0 \\
HCG14B    & 2001/01/01 & 3x1200 & 0 \\
HCG16A    & 2000/12/31 & 3x1200 & 90 \\
HCG16B    & 2000/12/31 & 3x1200 & 90 \\       
HCG16X    & 2000/12/31 & 3x1200 & 90 \\
HCG22A    & 2000/12/30 & 3x1200 & 30  \\
HCG22B    & 2001/01/01 & 3x1200 & 330 \\
HCG22D    & 2000/12/30 & 3x1200 & 30  \\
HCG22E    & 2000/12/30 & 3x1200 & 30  \\
HCG22X    & 2001/01/01 & 3x1200 & 330 \\
HCG25B    & 2000/12/31 & 3x1200 & 0 \\
HCG25F    & 2000/12/31 & 3x1200 & 0 \\
HCG32A    & 2000/12/30 & 3x1200 & 330 \\
HCG32B    & 2000/12/30 & 3x1200 & 330 \\
HCG32D    & 2000/12/30 & 3x1200 & 330 \\
HCG40A    & 2000/12/31 & 3x1200 & 0 \\
HCG40D    & 2000/12/31 & 3x1200 & 0 \\
\hline
\end{tabular} 
\end{center} 
\caption{Observational parameters for both the 2000 and 2001 observing runs.
HCG16X is a previously unidentified galaxy projected close to HCG16D at
$\alpha$=02$^h$09$^m$46.6$^s$, $\delta$=-10$^o$11'00.7'' (J2000). Its membership of HCG16 
is confirmed by its recession
velocity (see Table \ref{gal_props}). HCG22X has a 2MASS identification of
J03032308-1539079. The ZM identification refers to Zabludoff \&
Mulchaey (1998).}\label{obs}
\normalsize  
\end{table} 

Spectra were obtained on the ESO NTT telescope on 2000 April 9-11th and 
2000 Dec. 30-31st -- 2001 Jan. 1st (hereafter 2001 Jan run). 
Table 1 summarises the observational parameters. Seeing was around 1 
arcsec for both observing runs. The observations consist of a combination 
of long-slit and multi-object spectrograph (MOS) spectra obtained using the 
red arm of the EMMI instrument. All MOS masks had a slit-width 1.87 arcsec.
Long-slit spectra (slit-width 2.0 arcsec) were taken of additional galaxies 
not covered by the MOS observations.  In general, slits were 
not aligned along a preferred axis of the galaxies. All spectra were taken 
with ESO Grism\# 5 and cover the wavelength range from $\sim$4000 to 6600 \AA. 
This wavelength range covers 25 Lick indices. However,
redshift and vignetting effects cause the loss of indices at the extremes
of this range. The spectral resolution was 6.4\AA ~FWHM.
Spectra of seven Lick standard stars, as well
as spectrophotometric standards, were taken during the runs.
Exposures of a He+Ar lamp, with an order sorting filter BG38, were taken
for wavelength calibrations. Dome and sky flats were obtained for
flat-fielding.

Initial data reductions were carried out using standard tasks
within IRAF. Briefly, the data were bias-subtracted, trimmed,
bad pixels and cosmic rays removed, and flat-fielding carried out
using dome flats. Individual spectra were then extracted and sky-subtracted
using the {\it apall} task. Sky-subtractions in MOS data are notoriously
difficult, and despite careful selection of sky regions, sky-line residuals
were evident in some MOS spectra after sky-subtraction. The method of
analysis of these spectra was therefore designed to detect and ameliorate the 
effects
of such residuals (see Section \ref{agez}). Spectra were extinction corrected 
and flux calibrated using flux standard stars observed over several nights. 
For each galaxy we obtained three spectra of the same exposure time. These 
were co-added before extraction of the central regions. The extraction 
aperture was 10 pixels (2.7$^{''}$) wide for both the long-slit and MOS 
spectra. 

\begin{figure}
\centerline{\psfig{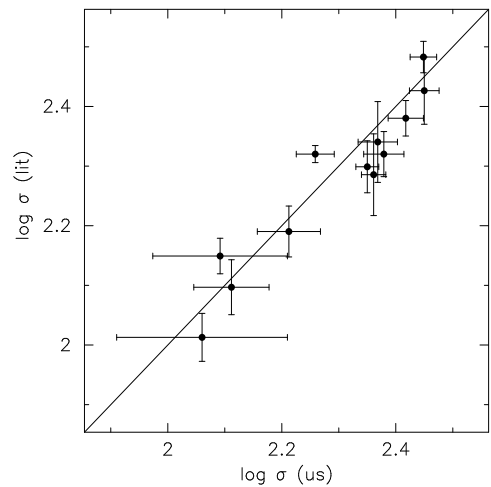}}
\caption{\label{comp_sig} Velocity dispersion estimates are compared to values from
Hypercat (with the exception of HCG32A, which was
taken from de la Rosa, Carvalho \& Zepf 2001a). The one-to-one
line is also shown. Good agreement with literature values is achieved.}
\end{figure}

Determination of recession velocity and velocity dispersion in galaxies 
was carried out using the IRAF task {\it fxcor}. The results of these 
determinations are given in Table \ref{gal_props}. A comparison of velocity 
dispersion measurements with values published in Hypercat
(http://www-obs.univ-lyon1.fr/hypercat/) are shown in 
Fig. \ref{comp_sig}. In this plot the data given in Table \ref{gal_props} have 
been aperture corrected to an effective aperture size of 1.19 {\it h$^{-1}$} 
kpc according to J{\o}rgensen \etal (1995). The effective aperture is approximated in 
J{\o}rgensen \etal (1995) by D=2.05$\times$(xy/$\pi$)$^{1/2}$. This 
yields effective apertures of 2.6$^{''}$ for the MOS data and 2.7$^{''}$ for 
the long-slit data. The average value (2.65$^{''}$) was used for aperture 
corrections. The estimated extent of these apertures (in kiloparsec) are given 
in Table \ref{gal_props}. Fig. \ref{comp_sig} shows that comparisons between 
aperture corrected values of log $\sigma$ from this study and those given in 
Hypercat are generally within errors.

\begin{table*} 
\footnotesize 
\begin{center} 
\begin{tabular}{lcccrcrcc} 
\hline 
   Name & Run    &Type &   B   & Velocity & M$_{B}$&$\sigma$~~~~& Aperture\\
        &     &     & (mag) & (km/s)   & (mag)  & (km/s) &  (kpc)  \\
\hline
{\bf Loose Group}& & & & & & & \\
NGC~3305 (LGG211)&2001 Jan LS & E0 & 13.77 & 3982 & --19.9 & 242(16) & 0.68\\
NGC~5046        & 2000 Apr LS & E  & 13.68 & 2209 & --18.7 & 135(24) & 0.38\\
NGC~5049        & 2000 Apr LS & S0 & 14.75 & 2958 & --18.2 & 122(18) & 0.51\\
\hline
{\bf Field}& & & & & &  & \\
NGC~2502        &2001 Jan LS  & SB0& 13.26 & 1093 & --17.6 & 169(36) & 0.19\\
NGC~3203        &2001 Jan LS  & S0 & 13.10 & 2503 & --19.4 & 128(21) & 0.43\\
NGC~6684        &2000 Apr LS  & SB0& 11.31 &  846 & --19.0 & 119(21) & 0.14\\
\hline
{\bf Compact Group}& & & & & & & \\
HCG4C          &2001 Jan MOS & E  & 16.08 &18306 & --20.9 & 182(19) & 3.14\\
HCG16X         &2001 Jan MOS & -- & --    & 3986 &    -- &  76(13) & 0.68\\
HCG22A         &2001 Jan MOS & E  & 12.85 & 2662 & --19.9 & 188(22) & 0.46\\
HCG22D         &2001 Jan MOS & SB0& 15.28 & 9268 & --20.2 & 236(16) & 1.59\\
HCG22E         &2001 Jan MOS & E  & 15.85 & 9579 & --19.7 & 168(19) & 1.64\\
HCG22X         &2001 Jan MOS & -- &  --   & 9244 &   --  & 242(20) & 1.58 \\
HCG25F         &2001 Jan MOS & S0 & 16.82 & 6328 & --17.8 & 123(25) & 1.08\\
HCG32A         &2001 Jan MOS & E  & 14.93 &12422 & --21.2 & 238(13) & 2.13\\
HCG32B         &2001 Jan MOS & S0 & 16.11 &12305 & --20.0 & 219(15) & 2.11\\
HCG32D         &2001 Jan MOS & S0 & 16.51 &12247 & --19.6 & 231(11) & 2.10\\
HCG40A         &2001 Jan MOS & E  & 13.77 & 6604 & --21.0 & 232( 6) & 1.13 \\
HCG40B         &2001 Jan LS  & S0 & 15.02 & 6690 & --19.7 & 217(22) & 1.15\\
HCG42A         &2000 Apr MOS & E  & 12.08 & 3792 & --21.4 & 291(30) & 0.65\\
HCG42C         &2000 Apr MOS & E  & 14.24 & 4045 & --19.4 & 157(34) & 0.69\\
HCG62:ZM19     &2000 Apr LS&E& 15.27 & 4204 & --18.5 & 209(22) & 0.72\\
HCG86A         &2000 Apr MOS & E  & 13.97 & 5955 & --20.5 & 292(12) & 1.02\\
HCG86B         &2000 Apr MOS & E  & 14.80 & 5817 & --19.6 & 248(17) & 1.00\\
\hline
{\bf Spiral Bulges}& & & & & & & \\
HCG4A        &2001 Jan MOS   & SBbc&14.16 & 8042 & --21.0 &  83(20) & 1.38\\
HCG14A       &2001 Jan MOS   & Sa & 14.44 & 5977 & --20.1 &  93(11) & 1.02 \\
HCG14B       &2001 Jan MOS   & Sa & 14.77 & 5386 & --19.5 & 134(29) & 0.92\\
HCG16A       &2001 Jan MOS   & Sb & 12.91 & 4032 & --20.7 & 147(50) & 0.69 \\
HCG16B       &2001 Jan MOS   & Sa & 13.73 & 3876 & --19,8 & 171(32) & 0.67\\
HCG22B       &2001 Jan MOS   & Sa & 15.47 & 2663 & --17.3 &  76(31) & 0.46\\
HCG25B       &2001 Jan MOS   & SBa& 15.21 & 6334 & --19.4 & 134(18) & 1.09\\
HCG40D       &2001 Jan MOS   & S0a& 15.06 & 6760 & --19.7 & 141(25) & 1.16\\
HCG62:ZM22   &2001 Jan LS& Sc&15.36& 4843 & --18.7 & 114(16) & 0.83\\ 
\hline
\end{tabular} 
\end{center} 
\caption{Galaxy properties for both observing runs. Galaxy types are averages 
of values given in Hypercat. Total B magnitudes are from Paturel \etal (2000) if
available and from NED if not. Absolute magnitudes are calculated 
from the estimated recession velocity assuming H$_0$ = 75 km/s/Mpc. 
Recession velocities and
velocity dispersions are estimated from our spectra. Velocity dispersion
errors are given in brackets. Recession
velocities are \emph{not} heliocentric velocities but rather \emph{observed}
velocities.  
The effective aperture size is calculated according to J{\o}rgensen \etal (1995).}
\label{gal_props}
\normalsize  
\end{table*} 

\begin{figure}
\centerline{\psfig{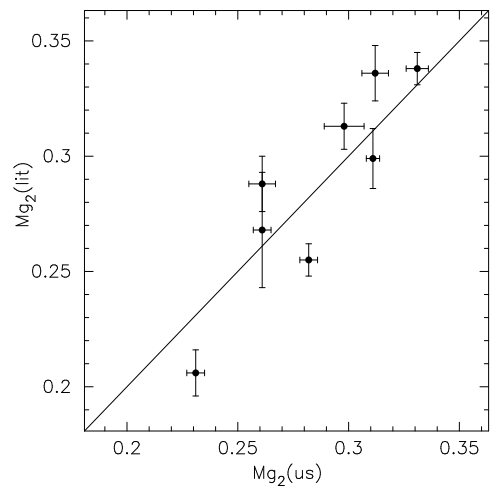}}
\caption{\label{Mg2comp} Comparison of Mg$_2$ values from Hypercat
with those measured here (aperture corrected).  The one-to-one line is also
shown. Reasonable agreement with literature values is achieved.}
\end{figure}

\section{Measurement of Lick indices}
\label{ind_meas}
Raw Lick indices were measured using our own code (see PS02). The code
measures indices at the wavelength dependent resolution detailed in Worthey
\& Ottaviani (1997). The measured indices at this point still
require correcting for both velocity dispersion and differences in flux
calibration between our study and
the Lick system.

\subsection{Calibration to the Lick system}
The procedure used for correcting Lick indices for galaxy velocity dispersion
was the same as that given in
PS02. Briefly, for galaxies in which the velocity dispersion, when combined 
in quadrature with the instrumental broadening, resulted in a resolution 
\emph{higher} than that of the Lick system, the galaxy spectra 
were convolved with an appropriate Gaussian prior to index measurement. 
This procedure is clearly not applicable to galaxies in which the combined 
resolution exceeded that of Lick system. Therefore, indices were measured in
7 Lick standard stars after convolving the spectra  with a series of Gaussians 
of known widths. This permits estimation of correction factors for indices in 
these galaxies. The method of calculating the correction factors was the same 
as detailed in PS02. The results from the present study
were almost identical to those of PS02, as well as those given in 
Kuntschner (2000).

The final correction required in order to fully calibrate to the Lick system is 
to allow for differences in flux calibration. To estimate this correction, 
Lick indices were measured 
in each of the observed 7 Lick standard stars. Stars were broadened to 
the appropriate Lick resolution prior to index measurement (see PS02). The 
difference between our measurements of the Lick standard 
stars and those taken on the Lick/IDS system (Worthey \etal 1994) are 
given in Table 3. Differences are generally smaller than the scatter and 
typical rms error per observation of the Lick calibrators (Worthey 
\etal 1994). Nonetheless, these offsets were applied to the velocity
dispersion corrected indices and the error in the mean added in quadrature
to index errors for the purposes of SSP fitting (Section \ref{agez}).

\begin{table} 
\footnotesize 
\begin{center} 
\begin{tabular}{lcrcc} 
\hline 
Index       &Units & Lick - our value   & Error   & Lick rms\\
            &      &                    & in mean & per obs.\\
\hline
H$\delta_A$ & \AA  &  0.229 $\pm$ 0.599 &  0.226  & 0.64\\ 
H$\delta_F$ & \AA  & -0.153 $\pm$ 0.429 &  0.162  & 0.40\\
CN$_1$	    & mag  & -0.014 $\pm$ 0.041 &  0.013  & 0.02\\
CN$_2$	    & mag  & -0.012 $\pm$ 0.012 &  0.004  & 0.02\\
Ca4227	    & \AA  & -0.089 $\pm$ 0.171 &  0.060  & 0.27\\
G4300	    & \AA  & -0.489 $\pm$ 0.251 &  0.079  & 0.39\\
H$\gamma_A$ & \AA  & -0.253 $\pm$ 0.432 &  0.153  & 0.48\\
H$\gamma_F$ & \AA  & -0.118 $\pm$ 0.183 &  0.064  & 0.33\\
Fe4383	    & \AA  & -0.015 $\pm$ 0.542 &  0.192  & 0.53\\
Ca4455	    & \AA  &  0.070 $\pm$ 0.437 &  0.155  & 0.25\\
Fe4531	    & \AA  & -0.382 $\pm$ 0.258 &  0.091  & 0.42\\
C4668	    & \AA  & -0.839 $\pm$ 0.270 &  0.095  & 0.64\\
H$\beta$    & \AA  & -0.052 $\pm$ 0.153 &  0.048  & 0.22\\
Fe5015	    & \AA  & -0.104 $\pm$ 0.295 &  0.104  & 0.46\\
Mg$_1$	    & mag  &  0.016 $\pm$ 0.012 &  0.004  & 0.01\\
Mg$_2$	    & mag  &  0.023 $\pm$ 0.011 &  0.003  & 0.01\\
Mgb	    & \AA  &  0.024 $\pm$ 0.135 &  0.043  & 0.23\\
Fe5270	    & \AA  & -0.324 $\pm$ 0.286 &  0.090  & 0.28\\
Fe5335      & \AA  & -0.394 $\pm$ 0.243 &  0.077  & 0.26\\
Fe5406	    & \AA  & -0.181 $\pm$ 0.127 &  0.045  & 0.20\\
Fe5709	    & \AA  & -0.030 $\pm$ 0.147 &  0.052  & 0.18\\
Fe5782	    & \AA  &  0.045 $\pm$ 0.126 &  0.045  & 0.20\\
Na D	    & \AA  & -0.097 $\pm$ 0.200 &  0.063  & 0.24\\
TiO$_1$	    & mag  &  0.003 $\pm$ 0.007 &  0.002  & 0.01\\
TiO$_2$     & mag  & -0.008 $\pm$ 0.005 &  0.002  & 0.01\\
\hline
\end{tabular} 
\end{center} 
\caption{Offsets from our index measurements to those of the Lick system. 
Mean offset and scatter about the mean for 6 Lick standard stars observed 
during both 2000 and 2001 runs are given. Error in the mean is also given.
This error must be added in quadrature to errors in Tables
\ref{indices1} to \ref{indices4} prior to age/metallicity determinations.
Also given is the typical rms error 
per observation of the Lick calibrators (see Worthey \etal 1994).}
\label{c2l}
\normalsize  
\end{table}

Comparison of Mg$_2$ index measurements with values published in Hypercat are shown
in Fig. \ref{Mg2comp}. Data in this plot are all aperture corrected according 
to J{\o}rgensen (1997). Only Mg$_2$ is compared as no other indices have been 
reported in previous studies that include HCGs common to this work. Agreement 
can be seen to be reasonably good, although the scatter is slightly larger than the errors. 
The cause of this scatter is unknown. However, it 
should be noted that Mg$_2$ is one of $\sim$20 indices used in
the derivations of age and metallicity detailed in Section \ref{agez}.
Consequently, the effects on this (or any other) individual index 
on the  final age/metallicity determinations is expected to be small.

\begin{figure*}
\centerline{\psfig{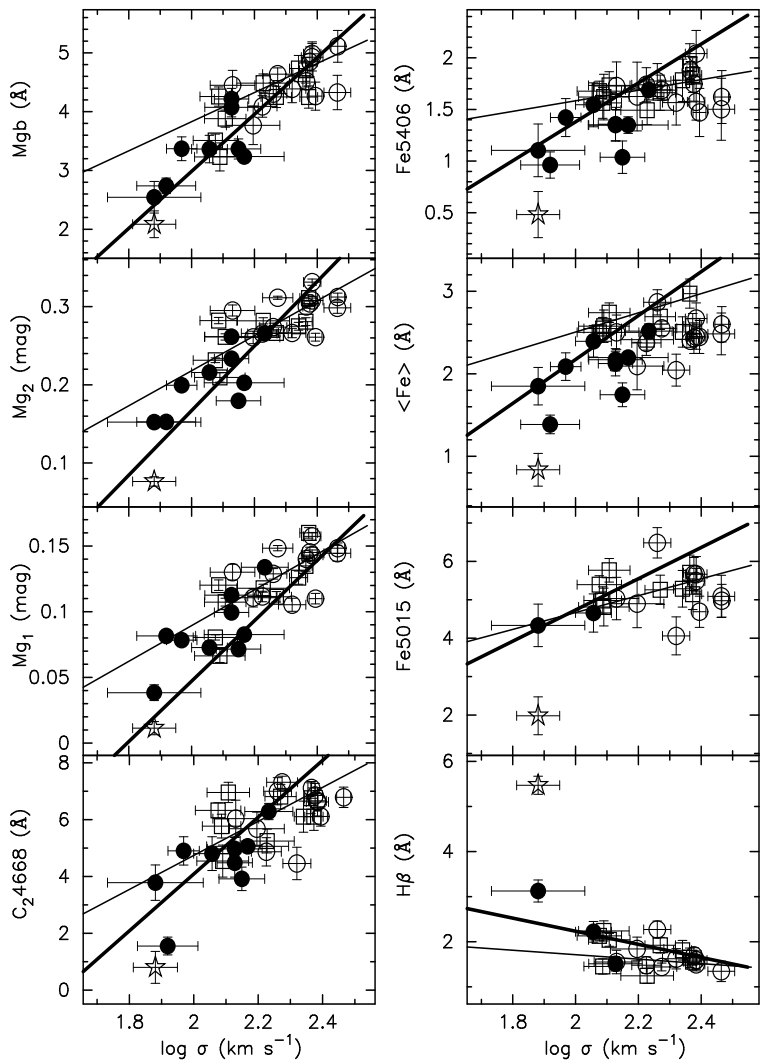}}
\caption{\label{index_sig}
Index--velocity dispersion relations are shown. Elliptical galaxies are shown by open
circles, S0s by open squares and spiral bulges by filled circles.
The starburst galaxy (HCG16X) is marked by a star symbol.
Data have been aperture corrected according to the prescription in
Appendix A of J{\o}rgensen (1997). The thin line in the Mg$_2$ plot
shows the relation for elliptical and S0 galaxies from Bernardi 
\etal (1998). This samples includes 931 field, group and cluster galaxies. Thin lines 
in the remaining plots are the Fornax E/S0 relations 
from Kuntschner (2000) (after conversion of line indices to a linear 
index scale and aperture correction). Thick lines in all plots are the 
PS02 relations for spiral galaxy bulges (after aperture correction). 
Although not identified separately, Arp-Madore galaxies and those in 
the NGC~5044 loose group follow the same general trends as HCG galaxies.}
\end{figure*}

\subsection{Correlations with velocity dispersion}
The trends of various indices with velocity dispersion are shown in Fig.
\ref{index_sig}. The thin line shown in the Mg$_2$ plot is the  correlation
reported by Bernardi et al (1998). Thin lines in the remaining plots are
from the data of Kuntschner (2000) for Fornax cluster galaxies (after
conversion of line indices to a linear index scale and aperture correction).
The thick lines are relations  for spiral galaxy bulges from PS02 (after aperture correction).
While there are small off-sets evident in some indices (most noticeably $<$Fe$>$),
agreement with the previously published correlations are generally good.

\begin{figure}
\centerline{\psfig{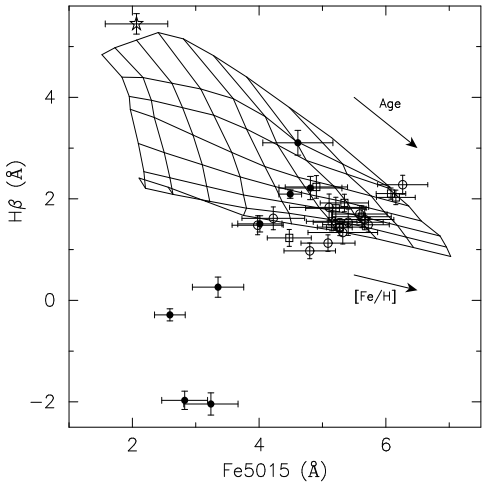}}
\caption{\label{Fe5015_Hbeta} Plot of H$\beta$ against Fe5015. Symbols as in
Fig. \ref{index_sig}. Grid lines represent lines of constant [Fe/H] (from -1.5
to +0.5 dex on 0.25 dex steps) and age (1, 1.5, 2, 3, 5, 8, 12 and 17 Gyr; bottom
line) for solar abundance ratio SSPs. The plot shows the
effects of emission on these indices in some galaxies i.e. they lie below
the grids. One galaxy (HCG4A)
with both Fe5015 and H$\beta$ of $\sim$ --5 $\AA$  has been omitted from this plot.}
\end{figure}

\subsection{Emission}
A plot of H$\beta$ against Fe5015 is shown in Fig. \ref{Fe5015_Hbeta}. 
Both of these indices are susceptible to emission. It is
clear from this diagram that a number of galaxies fall below the grids.
This is consistent with the presence of emission in these galaxies. However,
for galaxies lying just below the grids it \emph{might} reflect an overestimation of
H$\beta$ in the SSPs. Unfortunately, it was 
not deemed possible to estimate emission by the procedure of measuring 
[OIII]5007 (Gonzalez 1993) due to the relatively low signal-to-noise 
of the data and the small number of suitable template stars observed. The
identification and handling of these emission affected indices was therefore
carried out by consideration of \emph{all} indices in a 
galaxy. This procedure is described in the Section \ref{agez}.

\section{Determinations of age, metallicity and abundance ratios}
\label{agez}
This section details the estimation of age, [Fe/H] and [E/Fe], where [E/Fe] 
is the abundance ratio of enhanced elements; C,O, Mg etc to Fe (see PS02). 
To facilitate comparison of results to the studies of T00, and PS02, 
determinations 
of log(age), [Fe/H] and [E/Fe] (hereafter; derived parameters) were carried
out using Vazdekis (1999) single stellar populations (SSPs) and Tripicco 
\& Bell (1995) index sensitivities, as described in PS02\footnote{T00 in fact used Worthey 
(1994) SSPs and a slightly different method of estimating derived parameters.
However, these differences in combination result in values that differ 
by amounts similar to the quoted errors.}. Briefly, each index in the model 
SSPs was interpolated to give a grid of values for -1$<$log(age)$<$1.25 dex
and -1.675$<$[Fe/H]$<$0.5 dex in 0.025 dex steps. 
For each index, at each log(age), [Fe/H] step, values of the index were also
estimated for -0.3$<$[E/Fe]$<$0.6 dex, again in 0.025 dex steps. These
enhanced values were estimated by assuming an appropriate (55\%/44\%/3\%) mix of the three
stellar types whose index sensitivities were modeled by Tripicco
\& Bell (1995). The elements C, N, O, Mg, Na and Si were assumed to be in
the enhanced group (i.e. E in [E/Fe]), while Fe, Ca and Cr were taken to be
Fe peak elements. The fractional change in each index was then calculated by
assuming that all elements in the enhanced group vary by the same amount
when compared to the Fe peak elements. In this way a three dimensional grid
(with axes; log(age), [Fe/H] and [E/Fe]) was generated for each index.
The procedure used for estimating the 
derived parameters then involved simultaneously finding the best 
fit (by $\chi^2$ minimisation) of as many observed indices as possible ($\sim$20) to
the three dimensional grids of model indices. 

The rational for using a
large number of indices (rather than the 3 or 4 often used, e.g. Kuntschner
\etal 2001; Kuntschner \etal 2002; Caldwell, Rose \& Concannon 2003; Mehlert 
\etal 2003; Denicolo \etal 2003) was arrived at in the following way:
While each index is, to some degree, sensitive to each of the derived
parameters, the various degeneracies makes their estimation from individual
indices impossible. As well as errors in observations and reductions, we
must also consider the large number of uncertainties in the modeling
(comparison of complex populations to SSPs, errors in the SSP grids
themselves and the rather crude models of the effects of relative
abundances). However, despite these difficulties, it remains true that
each index does contain \emph{some} information on each of the derived
parameters. PS02 therefore postulated that, if a sufficiently large number 
of indices were
employed it may be possible to extract this information with acceptable
accuracy. In order to test
this hypothesis, PS02 used various combinations of indices to estimate
derived parameters, including estimates with all Balmer lines
excluded from the fitting procedure, i.e. \emph{using only metallicity
sensitive indices}. The more common combinations of 3 or 4 indices were
also tested. The estimates of derived parameters from the various
combinations of indices were shown to be in good agreement with only
moderate scatter ($\sim$ 0.1 dex) about values obtained using all indices (see
table 11 of PS02). PS02 therefore concluded that \emph{ages and metallicities can be
derived without the use of Balmer lines if a sufficiently large
number of indices are included in the fitting procedure}.

Given these considerations, the use of a large number of indices would seem
to have several distinct advantages;

\begin{itemize}

\item The procedure is particularly useful in spectra that are not emission
corrected, as affected indices can be identified and omitted from the 
determination (see below) with only a modest increase in errors. 

\item Derived parameters are less prone to reduction
errors, e.g. flux calibration error, stray cosmic rays, skyline residuals, 
velocity dispersion error, errors in conversion to the Lick system, etc.

\item Derived parameters are less prone to modeling errors in individual
indices.

\end{itemize}

While the process of estimating derived parameters used here is based on
that in PS02, there were some minor differences in methodology. Firstly, 
the molecular band indices were assumed to have an \emph{additive} Tripicco 
\& Bell (1995) correction (in line with Thomas et al. 2003). In addition, Ti 
was included amongst the enhanced elements for the application of Tripicco 
\& Bell (1995). These modifications to the PS02 procedure make only 
small differences to the derived parameters. More importantly, due to the 
presence of emission affected indices and the signal-to-noise 
of our data, 
certain indices were `clipped' from the fitting process using the following 
procedure. Initially the best fit age/[Fe/H]/[E/Fe] were estimated with 
(the emission affected) Balmer lines and Fe5015 excluded from the fitting 
procedure. Deviations of the observed Balmer lines and Fe5015 from the 
best fit SSP values were then considered, and indices whose 
observed value deviated by more than 3$\sigma$ from the best fit SSP value 
were deemed to be suffering from emission. These indices were then permanently 
excluded from the fitting process. A 3$\sigma$ offset in H$\delta_A$, 
H$\delta_F$, H$\gamma_A$, H$\gamma_F$ and H$\beta$ (with typical errors of 
$\sim$ 0.4, 0.3, 0.4, 0.3 and 0.2 $\AA$ respectively) corresponds to log(age) 
differences of $\sim$ 0.25, 0.45, 0.25, 0.35 and 0.5 dex. The clipping 
procedure therefore only excludes indices when derived parameter estimates 
obtained from the remaining indices differ significantly from that implied by 
the clipped index. 

Once emission affected indices had been eliminated, another 
fit to SSP values was obtained, this time using all remaining indices. 
Again indices were clipped, this time using a 5$\sigma$ limit. The 5$\sigma$ 
clipping process was iterated (if necessary) until a fit was obtained in which 
no index deviated from best fit SSP values by more than 5$\sigma$. In fact only 
one galaxy (HCG86A) required a second iteration. All but one of the 5 indices 
clipped at this stage were found (by visual inspection) to be affected by 
sky-line residuals. 

The final fit (with no indices deviating by more than 
5$\sigma$) was taken as giving the final values. Indices affected by
vignetting and those clipped from the fitting procedure have been omitted 
from Tables \ref{indices1} to \ref{indices4}. The values of the derived 
parameters are given in Table \ref{dervals} with subscript
\emph{raw}. This is to indicate that there a two further corrections that
must be applied before values are fully calibrated.

The first correction relates to local abundance trends, as the values of 
[E/Fe]$_{raw}$ given in Table \ref{dervals} are enhancements with
respect to the stellar calibrators used in the construction of the SSPs.
These stars were selected from the solar neighbourhood and therefore possess an
inherent [E/Fe] that varies with [Fe/H] (e.g. Edvardsson \etal 1993;
Gustafsson \etal 1999; Bensby, Felzing \& Lundstrom 2003). 

In order to understand the need for the local abundance ratio
pattern correction, consider a galaxy, the observed spectra of which
give derived parameters; [E/Fe]$_{raw}$=0.0 and [Fe/H]=--0.5. This is equivalent
to saying that the observed spectrum is best matched by the SSP with
[Fe/H]=--0.5, \emph{without} the need to adjust indices using the Tripicco \& Bell
(1995) abundance ratio sensitivities.

Now, the stellar library used to construct the SSP is composed of
stars in the solar neighbourhood, and therefore reflect the local abundance
ratio pattern. Thus the SSP at [Fe/H]=--0.5 is generated using stars with an
[E/Fe] of $\sim$+0.25 dex (see, for example, Bensby, Feltzing \& Lundstrom 2003).
Therefore, given that the galaxy spectrum is best matched by the
\emph{unadjusted} [Fe/H]=--0.5 SSP, we must conclude that the SSP and the galaxy have the 
same [E/Fe] of +0.25 dex.

This example can be generalised by considering the value [E/Fe]$_{raw}$ to
represent the offset between [E/Fe] in the galaxy and [E/Fe] in the best
fit SSP. Finally, it should be noted that solar neighbourhood stars (and 
consequently the SSPs) with [Fe/H]$>$--1.0 have [E/Fe] that varies with [Fe/H]. 
This variation (see, for example, Bensby, Feltzing \& Lundstrom 2003) can be 
approximated by;
   
\begin{equation}
[E/Fe] = -0.5[Fe/H].     \\
\label{eqna}
\end{equation}

The correction for the local abundance ratio pattern therefore
estimates the final value of [E/Fe] in galaxies with [Fe/H]$>$--1.0 using;

\begin{equation}
[E/Fe]_{final}=-0.5[Fe/H]+[E/Fe]_{raw}.
\label{eqnb}
\end{equation}

This correction is clearly only approximate. A more accurate analysis must 
account for the variety of behaviours of elements within the E group with
[Fe/H]. This issue will be addressed in future works.

In many previous works the correction outlined above is either omitted 
altogether (e.g. T00; Kuntschner \etal 2002) or 
applied only to galaxies with [Fe/H]$<$0 (e.g. PS02; Thomas \etal 2003).
However, recent studies of local stars have suggested that it should be applied 
to all galaxies with [Fe/H]$>$-1.0, including those with [Fe/H]$>$0 (Felzing \& Gustafsson
1998; Gustaffsson \etal 1999; Bensby \etal 2003; see Appendix B). This is the procedure 
adopted in this paper. The data of T00 and PS02 have also been corrected 
to make them consistent with the HCG analysis. A more complete justification 
and detailed account of this correction are given in Appendix \ref{LARC}.

The second correction that must be made to the raw values is
that for the varying size of apertures within and between studies. This 
correction was carried out by the procedure described in Appendix A1.2 of 
J{\o}rgensen (1997) using equation A2 and values $\alpha$=+0.2 for [Fe/H]; 
$\alpha$=0.0 for [E/Fe]; and $\alpha$=--0.04 for log(age). These values 
were estimated from gradients in derived parameters found in Proctor (2002).

The final derived parameters (corrected for both aperture and local enhancement 
trends) are given in Table \ref{dervals} with the
subscript \emph{corr}. It is these values upon which the following analysis
is based. Values of [E/H] (where [E/H]=[E/Fe]+[Fe/H]) are also given in
this table. [E/H] is used here, rather than the more commonly 
quoted [Z/H], as this aids in consideration of the correlations and temporal 
evolution of the two element groups of interest. However, the difference between 
[E/H] and [Z/H] is less than 0.04 dex for all galaxies in the sample.

\section{Results and discussion}
\label{roamd}

\begin{figure}
\centerline{\psfig{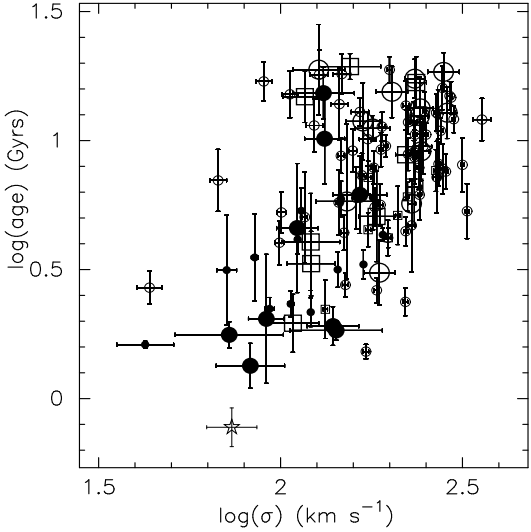}}
\caption{\label{age_sigma} Plots of log(age) with log $\sigma$ for 
galaxies in the current sample (large symbols) as well as T00
and PS02 samples (small symbols). Symbols are otherwise the same as Fig. 
\ref{index_sig}.}
\end{figure}

\begin{figure}
\centerline{\psfig{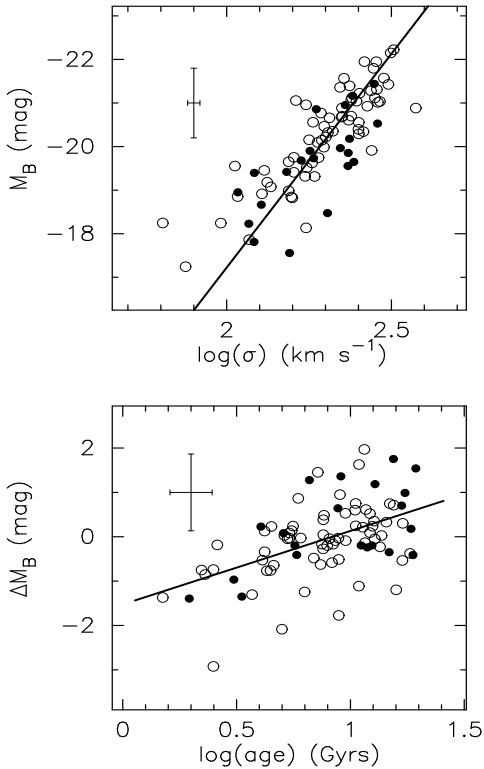}}
\caption{\label{fj} The top figure shows the Faber-Jackson relation (Faber
\& Jackson 1976) for
88 early-type galaxies in the three samples. Galaxies in the current work are
shown as solid symbols. The solid line is the relation from Forbes \& Ponman
(1999). Error bars represents an average error in log(age), while errors in
M$_B$ and $\Delta$M$_B$ are those which result from an assumed error of 350
km s$^{-1}$ (see text for details). Reasonable agreement is found with Forbes
\& Ponman relation. The bottom plot
shows residuals to the Faber-Jackson relation plotted against log(age).
The line shows the least-squares fit to the data. This correlation has a
$>$99.99\% significance.}
\end{figure}

The results presented in this section are fully corrected to the Lick
system. This includes both aperture correction and correction for the local
abundance trends (see Appendix \ref{LARC}). Plots include the measurements from 
T00 and PS02 data sets. From the present study we show results for 17
early-type galaxies and 9 spiral bulges in HCGs and 6 early-type galaxies in groups
or the field. The T00 sample includes 49
early-type galaxies of which 11 are in the Fornax cluster, 8 are in the Virgo 
cluster and 30 are in low density environments (groups and field).
NGC~221 and NGC~224 have been omitted from the original T00 sample due to the 
large aperture corrections required for such nearby galaxies. The PS02 sample
includes 6 Virgo cluster early-type galaxies, 2 galaxies in HGC68 (NGC~5353
and NGC~5354), 9 early-type galaxies in loose groups and the field as well as 
15 spiral bulges. The combined sample therefore contains 89 early-type
galaxies and 24 spiral bulges, in a variety of environments.

\subsection{Correlations for early-type galaxies}
A plot of luminosity-weighted age against velocity dispersion for the combined 
data sets (Fig. \ref{age_sigma}) indicates that the galaxies in the three samples exhibit a
large range of ages. The plot also indicates a trend between age and
velocity dispersion for both E/S0s and spiral bulges. Such a trend for
early-type galaxies is also evident in other studies (e.g. Caldwell \etal 
2003; Denicolo \etal 2003; Mehlert \etal 2003). However, some care must be 
taken when interpreting this figure as old, low velocity dispersion
(i.e. low luminosity) galaxies, which could populate the top left quadrant
of the plot may be excluded from any sample by signal-to-noise limitations.

The broad range of ages found in the E/S0 galaxies of these studies permits
testing of the Forbes \& Ponman (1999) finding of a correlation between
the residual of the Faber-Jackson relation and age. A plot of the Faber-Jackson
relation for the data in the two studies is shown in Fig. \ref{fj}. Values of M$_B$
were calculated using published distances to Virgo and Fornax clusters (16.8 and
18.4 Mpc respectively). However, for HCG and field galaxies, distances were 
calculated using simply recession velocities (a Hubble constant of 75 km s$^{-1}$ Mpc$^{-1}$ 
is assumed throughout). Fig. \ref{fj} (upper) shows that the relation 
given in Forbes \& Ponman (1999) is reproduced with a rms scatter $\sim$ 
0.8 mag. This scatter is equivalent to an uncertainty in distance estimates 
characterised by a recession velocity of only 350 km s$^{-1}$, well within
the random velocities of the field and group galaxies with respect to 
the Hubble flow. In the plot of residuals to the Faber-Jackson relation 
against age (lower plot) a correlation is
also found with confidence $>$99.99\%. This supports the Forbes \& Ponman (1999) 
finding and gives us confidence that 
the age/metallicity degeneracy has indeed been broken in these studies.

\begin{figure*}
\centerline{\psfig{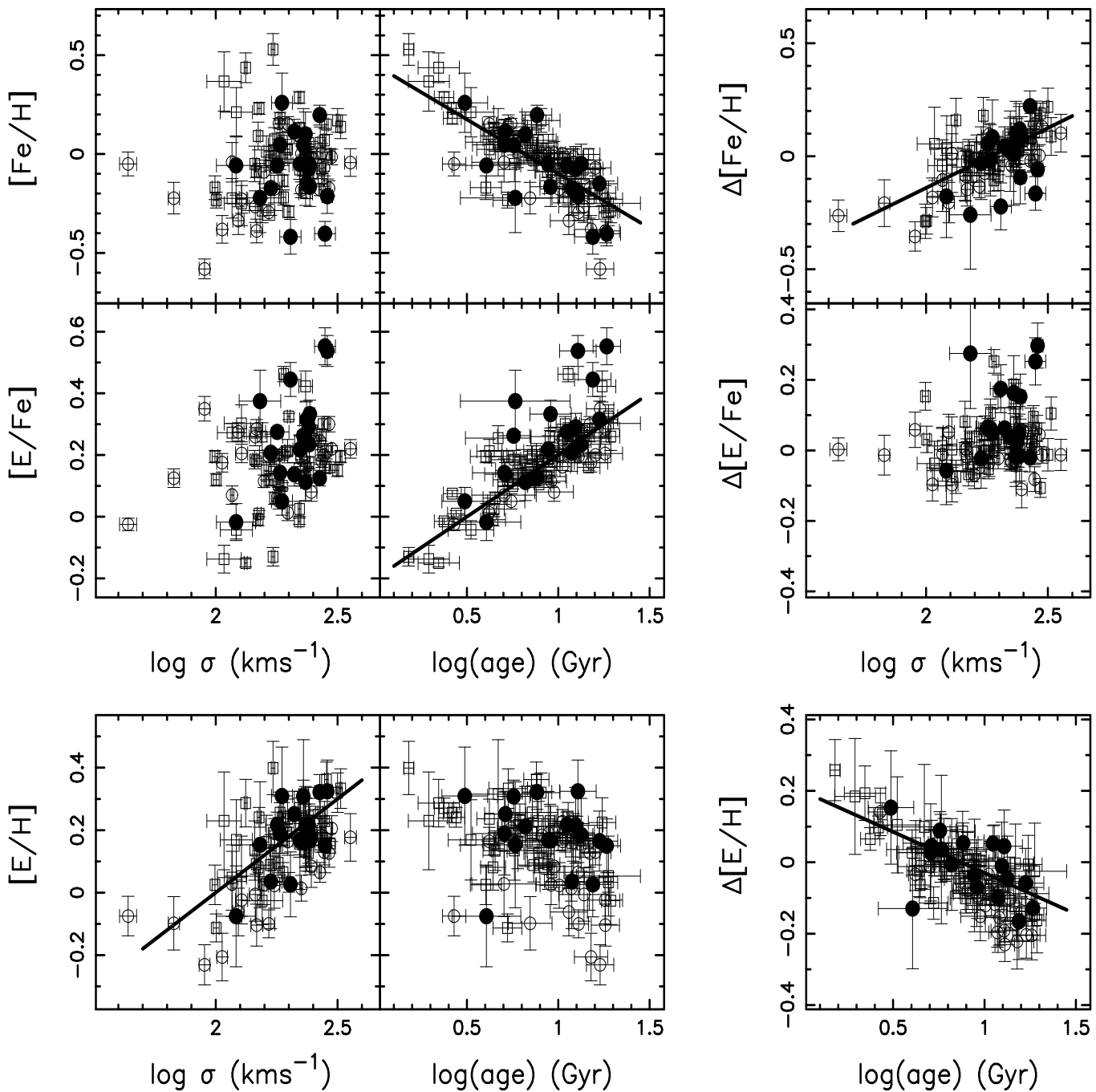}}
\caption{\label{E_agezsig} Plot of fully corrected derived parameters against
log $\sigma$ and log(age) (left-hand plots). Data from this study, T00 and PS02 are shown.
HCG galaxies are shown as filled circles, while cluster galaxies are shown
as open circles and those in loose groups and the field as open
squares. Lines are the correlations identified in the text. Strong
correlations are evident in both [Fe/H] and [E/Fe] with log(age), while, for
[E/H] the strongest correlation is with log $\sigma$. Residuals to these
correlations are shown on the right.
For [Fe/H] and [E/H] weak correlations also exist with log $\sigma$ and
log(age) respectively.}
\end{figure*}

Fig. \ref{E_agezsig} shows [Fe/H] and [E/Fe] (and their sum [E/H]) for
early-type galaxies in the three studies, plotted against velocity dispersion and
age. Bearing in mind that 89 galaxies are plotted in each of these figures,
strong correlations of both [Fe/H] and [E/H] with age are evident. Fits were
obtained (by least-squares fitting) to each of these relations. These
are shown in Fig. \ref{E_agezsig} as
solid lines. The fits were obtained using the field/loose group galaxies only, 
as these galaxies; i) span the largest ranges in both age and log velocity
dispersion; ii) constitute over 50\% of the combined sample; and iii) this 
environment was sampled in all three studies. The correlations shown in Fig. 
\ref{E_agezsig} are consistent with the results of PS02 and T00 (once differences
in the applied correction for local abundance trends have been accounted for),
and are also evident in the HCG data alone. Thus we find age--metallicity
and age--abundance ratio relations which indicate that older galaxies are
more Fe poor and $\alpha$-element enhanced that younger galaxies.

For  [E/H], the strongest correlation is with velocity dispersion. This clearly 
indicates that while [Fe/H] and [E/Fe] are primarily dependent on age there is 
also some dependence on velocity dispersion. Fig. \ref{E_agezsig} also shows the 
residuals of fits to the correlations identified above (estimated by least squares 
fitting). It is evident from these plots that residuals to the [Fe/H]--log(age) 
relation correlate with log $\sigma$, while residuals to the [E/H]--log $\sigma$
relation correlate with log(age). There is also a suggestion that the residuals 
to the [E/Fe]--log(age) relation correlate with log $\sigma$, but this correlation 
is not statistically significant. It is also clear from the residuals in Fig. 
\ref{E_agezsig} that HCGs generally follow the field galaxy correlations
reasonably well, albeit with a tendency for some HCG galaxies to exhibit lower
[Fe/H] residuals and higher [E/Fe] residuals than both field and cluster galaxies.
However, these galaxies exhibit some of the highest $\alpha$-enhancements in the 
combined sample. The offsets in residuals in these galaxies may then simply 
indicate  limitations in the modeling process rather than real differences. It
therefore seems unwise to draw any conclusions from these galaxies.

Given the dependence of [Fe/H], [E/Fe] and [E/H] on both age and velocity
dispersion, and the similarity of the galaxies in all environments, full three 
dimensional fits of the form  [Fe/H]~=~$\alpha$log(age)+$\beta$log($\sigma$)+$\gamma$ 
were found to the whole galaxy sample (see also T00). These fits gave;\\

\begin{equation}
[Fe/H] = -0.60log(age)+0.65log \sigma-1.00     \\
\label{eqn1}
\end{equation}

\begin{equation}
[E/Fe] = ~~0.30log(age)+0.05log \sigma-0.20     \\
\label{eqn2}
\end{equation} 

\begin{equation}
[E/H] = -0.25log(age)+0.60log \sigma-1.00     \\
\label{eqn3}
\end{equation}

Equations \ref{eqn1} to \ref{eqn3} indicate that the gradients of [Fe/H]
and [E/H] with age differ significantly, with the rate of Fe evolution
approximately twice that of elements in the enhanced group. On the other hand, it is 
evident that the gradients of [Fe/H] and  [E/H] with velocity dispersion are
almost identical. This is a result of finding only a statistically insignificant correlation
between [E/Fe] and velocity dispersion. This result is in stark contrast to the
findings of many previous studies, including Trager \etal (2000) and PS02, the data
from which have been re-analysed here. The discrepancy
between the findings of this work and the previous studies is the
result of differences in the local abundance ratio correction applied 
in each study (see Appendix \ref{LARC}). This emphasises the importance of this
(often overlooked) correction. The lack of correlation between [E/Fe] and
velocity dispersion is interesting in the context of galaxy formation models, as the
`semi-cosmological' N-body-SPH models of Kawata (Kawata 2001; Kawata \&
Gibson 2003), which model the hierarchical formation of individual galaxies, 
fail to reproduce a correlation between [E/Fe] and velocity dispersion, while
the N-body-SPH monolithic collapse models of Pipino \& Matteuchi (2003) require 
gas infall timescales that anti-correlate with galaxy mass in order to reproduce 
the assumed correlation.  This again underlines the importance of the 
local abundance ratio corrections required to fully calibrate the derived
parameters.

In order to investigate possible variations in the derived parameters with 
environment we divided the
sample into four categories; clusters, massive groups, HCGs and finally small
groups/field. Cluster galaxies are those in Virgo and Fornax from T00 and
PS02. Massive group galaxies include those in the NGC~5044 group and all
other galaxies identified as having at least 10 neighbours in Garcia (1993).
All remaining galaxies not in HCGs were attributed to the small group/field
category. Galaxies with log $\sigma$ $\le$ 2.075 were omitted from this analysis 
to ensure that the mass distribution in the different environments were as 
similar as possible. Averages of log $\sigma$ and the derived parameters for 
galaxies in each environment are presented in Table \ref{envs}. Errors in the 
mean are also presented. The mean velocity dispersion and its range can be 
seen to be almost identical in the four environments. Table \ref{envs} also 
indicates that galaxies in low density environments (small groups/field) exhibit 
a lower average age (5.9 Gyr) than those in clusters (11.0 Gyr) and massive 
groups (10.2 Gyr). This is consistent with the results of previous authors 
(e.g. Rose \etal 1994; Kuntschner \etal 2002; Terlevich \& Forbes 2002). It is 
also qualitatively consistent with the predictions of hierarchical formation 
models (e.g. Baugh, Cole \& Frenk 1996; Kauffmann \& Charlot 1998). 

Galaxies in low density environments also appear to possess higher [Fe/H] than 
their cluster and massive group counterparts. This is again consistent with
the findings of Rose \etal (1994). In order to test the effect of differences in the
morphological mix (elliptical to S0) in each of the samples, average values
of derived parameters were also calculated for elliptical galaxies only. No
significant qualitative differences in the results shown in Table \ref{envs} 
were observed. 

\begin{table*}
\footnotesize
\begin{center}
\begin{tabular}{lccccc}
\hline
Environment        & N & Average    & Average    & Average     & Average \\
                  &    &log $\sigma$& log(age)   &  [Fe/H]     & [E/Fe]  \\
\hline		                    
HCG               & 18 & 2.32(0.02) & 0.93(0.05) & -0.07(0.04) & 0.26(0.04)\\
\hline		                    
Cluster           & 20 & 2.32(0.03) & 1.04(0.03) & -0.11(0.03) & 0.19(0.02)\\
Massive Group     & 13 & 2.34(0.04) & 1.01(0.05) & -0.07(0.03) & 0.25(0.03)\\
Small group/field & 30 & 2.28(0.02) & 0.77(0.05) & ~0.07(0.03) & 0.12(0.02)\\
\hline
\end{tabular}
\end{center}
\caption{The average values of derived parameters for E/S0 galaxies in the 
combined sample are given by environment. Errors in the mean are given in brackets.
Galaxies with log $\sigma<$2.075 are omitted to ensure
a close match in log $\sigma$ distributions in the four environments. HCG
galaxies can be seen to more closely resemble cluster/massive group galaxies
than those in small groups/field. No qualitative differences were observed
when only elliptical galaxies were considered.}
\label{envs}
\normalsize
\end{table*}

Comparison of HCG galaxies to galaxies in other environments (Table \ref{envs})
therefore shows them to possess distributions of derived parameters more in keeping
with those of galaxies in high density environments than those of galaxies in small
groups and the field, i.e they are on average older and possess higher
[Fe/H] than their low density environment equivalents. This is consistent with
the findings of de la Rosa \etal (2001b) who also found such
a difference in age and metallicity. However, the five
early-type galaxies that are the brightest in their respective compact groups
(i.e. HCG22A, HCG32A, HCG40A, HCG42A, HCG86A and HCG68A from PS02) have an average
log(age) of 1.03 $\pm$ 0.07 dex (10.7 $\pm$ 1.9 Gyr). This is \emph{older} than the
average of the HCG sample as a whole, with the remaining galaxies having an
average log(age) of 0.89 $\pm$ 0.07 dex (7.7 $\pm$ 1.4 Gyr). These findings contradict 
the prediction of the Kauffmann (1996) hierarchical formation models that the brightest 
galaxies in such groups should be the youngest. However, they are, at
least qualitatively, consistent with the Chiosi \& Carraro (2002) models of
monolithic collapse, although these models do not appear to be able to
explain the extremely young ages observed in some galaxies.

\subsection{Correlations for spiral bulges}
\begin{figure}
\centerline{\psfig{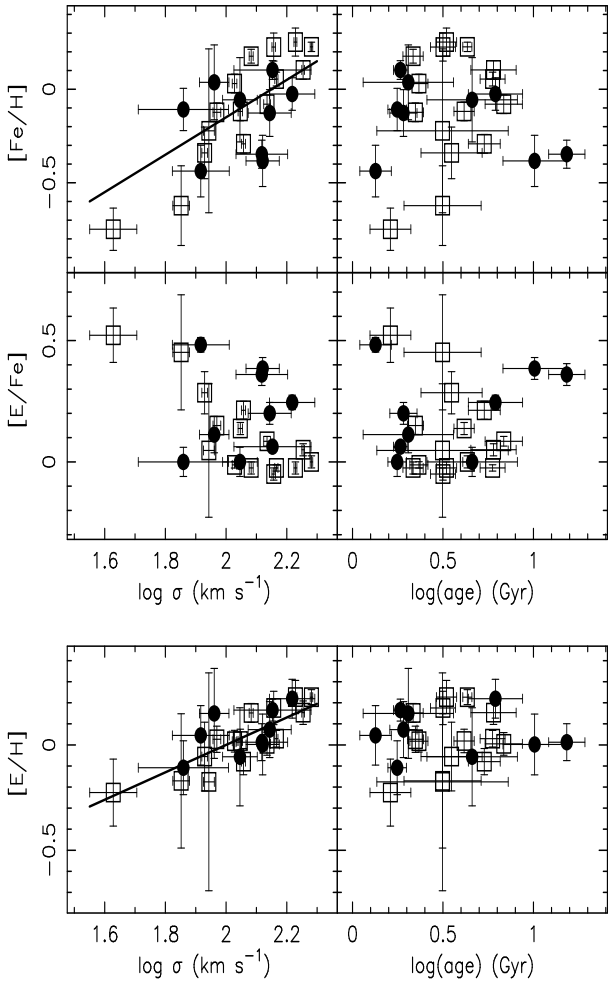}}
\caption{\label{B_agezsig} Plots of metallicity parameters in spiral bulges
against log(age) and log $\sigma$. Best fit lines (least squares fitting)
are shown for [Fe/H] and [E/H] against log $\sigma$. Spiral bulges in HCGs are
shown as filled circles, while those in all other environments are shown as
open squares.}
\end{figure}

The HCG data for spiral bulges are in good agreement with PS02 in that 
they exhibit lower average ages than early-type galaxies and show strong 
correlations of both [Fe/H] and [E/H] with log $\sigma$ (Fig. \ref{B_agezsig}). 
The gradient of the [E/H]--log $\sigma$ correlation in spiral bulges (0.65 dex/dex) 
is similar to that found in early-type galaxies (Equation \ref{eqn3}). 
Indeed, combined [E/H]--log $\sigma$ plots of spiral
bulges and early-type galaxies show a single overlapping relation (see Fig. 
\ref{larc2}). However, 
the gradient of the [Fe/H]--log $\sigma$ correlation in bulges (1.0 dex/dex) 
appears to differ from that in early-types (0.65 dex/dex). No correlations were 
found among the residuals to these correlations. Fig. \ref{B_agezsig} also 
indicates no significant differences in log(age), [Fe/H] or [E/Fe] between 
spiral bulges in HCGs and those in other environments.
We note that Proctor (2002) showed that, even in edge-on systems, disc
contamination was below 10\%, in the central regions of spiral bulges. Disc
contamination is therefore not considered significant in this study.

\section{Conclusions}
We have investigated the ages, metallicities and abundance ratios of galaxies 
in a variety of environments and compared them to those of Hickson Compact 
Group galaxies. It has been shown that early-type galaxies in all environments 
exhibit a range of ages. Residuals to the Faber-Jackson relation correlate 
reasonably well with age in all environments, consistent with
Forbes \& Ponman (1999) and giving confidence that the age/metallicity 
degeneracy has been broken.

Correlations were detected between age and velocity dispersion in both early-
and late-type galaxies, with high velocity dispersion galaxies exhibiting
older ages. This is, at leat qualitatively, consistent with previous studies
as well as the monolithic collapse models of Chiosi \& Carraro (2002). However,
studies must be extended to lower velocity dispersions to confirm this finding.

The distributions of age in the various environments (HCGs, clusters, 
massive groups and small groups/field) confirmed the prediction of hierarchical 
formation models (e.g. Baugh \etal 1996; Kauffmann \& Charlot 1998) and 
the findings of previous studies (e.g. Kuntschner \etal 2002;
Terlevich \& Forbes 2002) that galaxies in the field have a lower average age
than those in clusters. However, the finding that HCGs more closely resemble 
cluster galaxies than those in the field is inconsistent with current
hierarchical formation models.
The brightest galaxies in HCGs were also shown to be \emph{older} than their
less luminous companions. This is in stark contrast with the Kauffmann
(1996) prediction that the brightest galaxies in such 
groups should be younger than their less luminous neighbours.

Our study also confirmed the previously reported finding (de la
Rosa \etal 2001b) that early-type galaxies in HCGs are more metal poor 
than field galaxies. This again
suggests that the star formation histories of early-type galaxies in HCGs 
more closely resemble those of cluster galaxies than those in less dense 
environments such as small groups and the field.

Metallicities (given by [Fe/H] and [E/H]) were found to show strong
correlations with both age and velocity dispersion for early-type galaxies in all
environments. These correlations were shown to be consistent across all galaxy
environments. The correlations of [Fe/H]
and [E/H] with age were shown to differ significantly in that Fe abundance exhibits
a rate of change approximately twice that of the $\alpha$-elements in the
enhanced group. However, the [Fe/H] and [E/H] correlations
with velocity dispersion have similar gradients, as we find no significant correlation
between $\alpha$-element enhancement ration ([E/Fe]) and
velocity dispersion. We therefore conclude that this correlation, that has
been reported in many previous studies, is probably the
result of a failure to fully calibrate the derived parameters to the
Lick system. Specifically, the estimated values of [E/Fe] must be corrected
for the pattern of [E/Fe] with [Fe/H] found in the local stars used to
calibrate the Lick system. 

Spiral bulges were shown to exhibit the same strong correlations of [Fe/H] and
[E/H] with log $\sigma$ found in PS02. The [E/H]--log $\sigma$ relation in
bulges was shown to be the same as that in early-type galaxies. However, the
[Fe/H]--log $\sigma$ relations differ between early- and late-type galaxies.
No correlations with age were identified. No evidence was found for variation 
of star formation histories in spiral bulges with environment. We therefore
find that the star formation histories of the centres of spiral bulges
differ significantly from those in the centres of early-type galaxies.\\

\noindent{\bf Acknowledgments}\\
This work is based on observations made on the NTT, which is operated by ESO.
The authors acknowledge the data analysis facilities provided by IRAF, which
is distributed by the National Optical Astronomy Observatories and
operated by AURA, Inc., under cooperative agreement with the National
Science Foundation. R.P. thanks the Royal Society for funding that supported 
this work. \\

\noindent{\bf References}

\noindent
Athanassoula E., Makino J., Bosma A., 1997, MNRAS, 286, 825\\
Baugh C.M., Cole S., Frenk C.S., 1996, MNRAS, 283, 1361\\
Bensby T., Felzing S., Lundstr{\o}m I., 2003, A\&A, 410, 527\\
Bernardi M., Renzini A., da Costa L.N., Wegner G., Alonso M.V., 
Pellegrini P.S., Rité C., Willmer C.N.A., 1998, ApJ, 508, 143\\
Caldwell N., Rose J.A., Concannon K.D., 2003, AJ, 125, 2891\\
Chiosi C., Carraro G., 2002, MNRAS, 335, 335\\
Coziol R., Ribeiro A.L.B., de Carvalho R.R., Capelato H.V., 1998, ApJ, 493, 563\\ 
de la Rosa I.G., de Carvalho R.R., Zepf S.E., 2001a, AJ, 122, 93\\
de la Rosa I.G., Coziol R., de Carvalho R.R., Zepf S.E., 2001b, ApSS 276,717\\
Diaferio A., Gell M.J., Ramella M., 1994, AJ, 107, 868\\
Denicolo G., Terlevich R., Terlevich E., Forbes D., Terlevich A., 2003 in prep.\\
Edvardsson B., Andersen J., Gustafsson B., Lambert D.L., Nissen P.E.,
Tomkin J., 1993, A\&A, 275, 101\\
Felzing S., Gustafsson B., 1998, A\&AS, 129,237\\
Faber S.M., Jackson R.E., 1976, ApJ, 204, 668\\
Forbes, D., 1992, A\&AS, 92, 583\\
Forbes, D., Ponman, T., 1999, 309, 623\\
Garcia A.M., 1993, A\&AS, 100, 47\\
de la Rosa I.G., Coziol R., de Carvalho R.R., Zepf S., 2001, Ap\&SS,
276, 717\\
Gonz\'{a}lez J.J., 1993, Ph.D Thesis, University of California, Santa Cruz\\
Governato F., Tozzi P., Cavaliere A., 1996, ApJ, 458, 18\\
Gustafsson B., Karlsson T., Olsson E., Edvardsson B., Ryde N., 1999, A\&A,
342, 426\\
Hernquist, L., Katz, N., Weinberg, S., 1995, ApJ, 442, 57\\
Hickson  P., 1982, ApJ, 255, 382\\
Jones, L., Ponman, T., Forbes, D., 2000, 312, 319\\
J{\o}rgensen I., Franx M., Kj{\ae}rgaard P., 1995, MNRAS, 276, 1341\\
J{\o}rgensen I., 1997, MNRAS, 288, 161\\
Kauffmann G., 1996, MNRAS, 281, 487\\
Kauffmann G., Charlot S., 1998, MNRAS, 294, 705\\
Kawata D., 2001, ApJ, 558, 614\\
Kawata D., Gibson B.K., 2003, MNRAS, 340, 908\\
Kuntschner H., 2000, MNRAS, 315, 184\\
Kuntschner H., Lucey J.R., Smith R.J., Hudson M.J., Davies R.L., 2001, MNRAS, 
323, 615\\
Kuntschner H., Smith R.J., Colless M., Davies R.L., Kaldare R., Vazdekis A.,
2002, MNRAS, 337, 172
Mamon, G., 1986, ApJ, 307, 426\\
Mehlert D., Thomas D., Saglia R.P., Bender R., Wegner G., 2003, A\&A, 407, 423\\
Mendes de Olivera C., Hickson P., 1994, 427, 684\\
Nissen P.E., Edvardsson B., 1992, A\&A, 261, 255\\
Paturel G., Fang Y., Garnier R., Petit C., Rousseau J., 2000, A\&AS, 146, 19\\ 
Pipino A., Matteuci F., 2003, (in prep)
Poggianti B.M., Bridges T.J., Mobasher B., Carter D., Doi M.,
Iye M., Kashikawa N., Komiyama Y., Okamura S., Sekiguchi M.,
Shimasaku K., Yagi M., Yasuda N., 2001, ApJ, 562, 689\\
Ponman, T., Bourner, P., Ebeling, H., Bohringer, H., 1996, MNRAS,
283, 690\\
Proctor R.N., 2002, Ph.D Thesis, University of Central Lancashire, Preston, UK\\
(http://www.star.uclan.ac.uk/~rnp/research.htm)\\
Proctor R.N., Sansom A.E., 2002, MNRAS, 333, 517 ({\bf PS02})\\
Rose J.A., Bower R.G., Caldwell N., Ellis R.S., Sharples R.M., Teague P.,
1994, AJ, 108, 2054\\
Rubin, V., Hunter, D., Ford, W., 1991, ApJS, 76, 153\\
Terlevich A.I., Forbes D.A., 2002, MNRAS, 330, 547\\
Thomas D., Maraston C., Bender R., MNRAS, 339, 897\\
Tovmassian, H., Tiersch, H., Martinez, O., Yam, O., 2000, {\it
  Small Galaxy Groups}, ed. M. Valtonen \& C. Flynn, ASP Conference
  Series, Vol. 209, San Francisco\\
Trager, S., Faber, S., Worthey, G., Gonzalez, J., 2000, AJ, 120, 165 ({\bf T00})\\
Tripicco M.J., Bell R.A., AJ, 110, 3035 ({\bf TB95})
Vazdekis A., 1999, http://www.iac.es/galeria/vazdekis/
Verdes-Montenegro, L., Yun M.S., Perea J., del Olmo A., Ho P.T.P., 1998, ApJ, 497, 89\\
White, S., 1990, {\it Dynamics and Interactions of Galaxies},
ed. R. Wielen, Heidelberg, Springer, p 380\\
Worthey G., 1994, ApJS, 95,107
Worthey G., Ottaviani D.L., 1997, ApJS, 111, 377\\
Worthey G., Faber S.M., Gonz\'{a}lez J.J., Burstein D., 1994, ApJS, 94, 687\\
Zabludoff A.I., Mulchaey J.S., 1998, ApJ, 496, 39\\
Zepf, S., Whitmore, B., 1991, ApJ, 383, 542\\

\begin{appendix}
\section{Lick Indices}

\begin{table*}
\footnotesize 
\begin{center} 
\begin{tabular}{lccccccccccc} 
\hline
Galaxy & H$\delta_A$ & H$\delta_F$ & CN$_1$ & CN$_2$ & Ca4227 &
H$\gamma_A$ & H$\gamma_F$ & Fe4383 & Ca4455 & Fe4531 & C$_2$4668 \\
\hline
{\bf Loose Group}&&&&&&&&&&&\\                                
NGC~3305       	 &-3.121 & -0.259 & 0.106  & 0.142  & 1.247  & -6.886 & -2.250 & 5.039  & 1.106  & 3.251  & 7.131   \\ 
             	 &0.352  & 0.228  & 0.008  & 0.010  & 0.142  & 0.285  & 0.181  & 0.344  & 0.179  & 0.256  & 0.360   \\ 
NGC~5046       	 &-3.316 & -0.534 & 0.084  & 0.118  & 1.070  & -7.191 & -2.640 & 5.184  & 1.968  & 3.179  & 6.772   \\ 
               	 &0.655  & 0.429  & 0.015  & 0.018  & 0.261  & 0.523  & 0.337  & 0.607  & 0.296  & 0.447  & 0.648   \\ 
NGC~5049       	 &-3.264 & -0.251 & 0.073  & 0.106  & 1.045  & -6.776 & -2.065 & 5.188  & 1.857  & 3.309  & 6.325   \\ 
               	 &0.412  & 0.266  & 0.010  & 0.011  & 0.166  & 0.332  & 0.210  & 0.389  & 0.192  & 0.285  & 0.418   \\ 
\hline
{\bf Field}&&&&&&&&&&&\\
NGC~2502       	 &-2.508 & -0.412 & 0.079  & 0.109  & 1.108  & -7.984 & -2.732 & 6.243  & 1.385  & 4.074  & 6.226   \\ 
             	 &0.465  & 0.307  & 0.011  & 0.013  & 0.179  & 0.366  & 0.232  & 0.399  & 0.205  & 0.294  & 0.419   \\ 
NGC~3203       	 &-1.681 & 0.103  & 0.051  & 0.082  & 1.110  &        & -1.609 & 5.896  & 1.656  & 3.580  & 7.727   \\ 
             	 &0.345  & 0.228  & 0.008  & 0.010  & 0.141  &        & 0.177  & 0.329  & 0.170  & 0.250  & 0.356   \\ 
NGC~6684        	 &-1.740 & 0.251  & 0.049  & 0.077  & 1.079  &        &        & 5.664  & 1.344  & 3.277  & 7.660   \\ 
               	 &0.264  & 0.174  & 0.006  & 0.007  & 0.108  &        &        & 0.253  & 0.128  & 0.186  & 0.271   \\ 
\hline
{\bf Compact Group}&&&&&&&&&&&\\ 
HCG4C	         &-0.573 & 0.910  & 0.052  & 0.095  & 1.004  & -5.841 & -1.119 & 5.772  & 1.792  & 3.358  & 6.630  \\
                 &0.422  & 0.278  & 0.010  & 0.012  & 0.191  & 0.369  &  0.228 & 0.439  & 0.220  & 0.328  & 0.478  \\
HCG16X           &       &        &        &        &        &        &        &        & 0.321  & 1.214  & 0.858   \\ 
                 &       &        &        &        &        &        &        &        & 0.248  & 0.371  & 0.562   \\ 
HCG22A           &-2.905 & -0.319 & 0.105  & 0.144  & 1.244  & -6.904 & -2.098 & 5.691  & 1.362  & 3.177  & 8.084   \\ 
                 &0.232  & 0.152  & 0.005  & 0.006  & 0.092  & 0.180  & 0.114  & 0.201  & 0.104  & 0.151  & 0.215   \\ 
HCG22D           &       & -0.038 & 0.075  & 0.113  & 1.303  & -6.687 & -2.261 & 4.919  & 1.841  & 2.461  & 6.104   \\ 
                 &       & 0.322  & 0.012  & 0.014  & 0.203  & 0.392  & 0.251  & 0.467  & 0.234  & 0.439  & 0.479   \\ 
HCG22E           &       &        &        &        & 0.946  & -6.361 & -1.592 & 5.080  & 1.270  & 2.875  & 4.856   \\ 
                 &       &        &        &        & 0.200  & 0.383  & 0.239  & 0.453  & 0.230  & 0.332  & 0.498   \\ 
HCG22X           &-2.147 & -0.145 & 0.079  & 0.121  & 1.409  & -7.700 & -2.375 & 5.137  & 1.561  & 3.542  & 6.624   \\ 
                 &0.412  & 0.273  & 0.010  & 0.012  & 0.177  & 0.350  & 0.226  & 0.444  & 0.227  & 0.334  & 0.492   \\ 
HCG25F           &       &        & 0.006  & 0.042  & 0.941  & -5.105 &        & 4.415  & 1.217  & 2.853  & 4.684   \\ 
                 &       &        & 0.011  & 0.013  & 0.205  & 0.405  &        & 0.503  & 0.255  & 0.375  & 0.564   \\ 
HCG32A           &-2.550 & -0.238 & 0.079  & 0.118  & 1.054  & -6.868 & -2.026 & 5.670  & 1.676  & 3.268  & 6.692   \\ 
                 &0.379  & 0.250  & 0.010  & 0.011  & 0.160  & 0.326  & 0.206  & 0.386  & 0.198  & 0.296  & 0.422   \\ 
HCG32B           &-2.252 & 0.179  & 0.062  & 0.111  & 1.349  & -6.173 &        & 5.224  & 1.616  & 3.325  & 5.968  \\
                 &0.478  & 0.313  & 0.011  & 0.013  & 0.204  & 0.412  &        & 0.497  & 0.253  & 0.371  & 0.546  \\
HCG32D           &-1.419 & 0.285  & 0.046  & 0.082  & 1.281  & -6.123 & -1.331 & 5.705  & 1.675  & 3.056  & 6.604   \\
                 &0.466  & 0.311  & 0.011  & 0.013  & 0.205  & 0.393  & 0.257  & 0.498  & 0.248  & 0.367  & 0.543  \\
HCG40A           &-2.437 & -0.109 & 0.097  & 0.136  & 1.095  &        &        & 5.373  & 1.606  & 3.253  & 7.312  \\
                 &0.217  & 0.143  & 0.005  & 0.006  & 0.092  &        &        & 0.216  & 0.109  & 0.159  & 0.230  \\
HCG40B       	 &-1.911 & 0.269  & 0.060  & 0.094  & 1.295  & -5.271 & -1.209 & 4.965  & 1.728  & 3.321  & 6.992   \\ 
             	 &0.387  & 0.252  & 0.009  & 0.011  & 0.162  & 0.329  & 0.205  & 0.396  & 0.213  & 0.294  & 0.430   \\ 
HCG42A           &-2.090 & -0.193 & 0.083  & 0.113  & 0.859  & -6.301 & -1.767 & 4.993  & 1.296  & 2.885  & 9.308   \\ 
               	 &0.290  & 0.192  & 0.008  & 0.009  & 0.143  & 0.300  & 0.185  & 0.378  & 0.204  & 0.313  & 0.480   \\ 
HCG42C        	 &       &        &        &        &        &        &        & 4.265  & 1.256  & 3.021  & 6.072   \\ 
               	 &       &        &        &        &        &        &        & 0.437  & 0.231  & 0.361  & 0.584   \\ 
HCG62:ZM19       &-2.415 & -0.346 & 0.065  & 0.101  & 1.068  & -5.938 & -1.639 & 3.832  & 1.285  & 2.815  & 4.749   \\
               	 &0.509  & 0.340  & 0.012  & 0.014  & 0.217  & 0.429  & 0.270  & 0.532  & 0.266  & 0.388  & 0.570   \\ 
HCG86A         	 &-2.347 &        & 0.124  & 0.168  & 0.795  & -6.981 &        & 5.021  &        & 2.892  & 7.032   \\ 
               	 &0.215  &        & 0.005  & 0.006  & 0.111  &  0.219 &        & 0.281  &        & 0.227  & 0.359   \\ 
HCG86B    	 &       &        & 0.083  & 0.127  & 0.924  & -6.588 &        & 5.210  & 1.205  & 2.879  & 6.345   \\ 
               	 &       &        & 0.005  & 0.006  & 0.100  &  0.208 &        & 0.284  & 0.151  & 0.217  & 0.342   \\ 
\hline
\end{tabular}
\end{center}
\caption{Calibrated Lick indices. The data are \emph{not} aperture
corrected. Omitted indices are those lost to vignetting effects and those
excluded from the estimation of derived parameters (Section
\ref{agez}).}\label{indices1}
\normalsize
\end{table*}

The indices presented in Tables \ref{indices1} to \ref{indices4} are for the
galaxies in the present study. Indices are fully
corrected to the Lick system but \emph{not} aperture corrected. Indices
affected by vignetting effects and those clipped from the fitting procedure
(see Section \ref{agez}) are omitted. 

\begin{table*}
\footnotesize
\begin{center}
\begin{tabular}{lccccccccccc}
\hline
Galaxy           &H$\beta$& Fe5015 & Mg$_1$ & Mg$_2$ &  Mgb & Fe5270 & Fe5335 &  Fe5406 & Fe5709 & Fe5782 & [MgFe] \\
\hline
{\bf Loose Group}&&&&&&&&&&&\\
NGC~3305       	 & 1.490  & 5.723  & 0.172  & 0.346  & 5.198  & 2.665  & 2.445  & 1.641  & 0.989  & 0.788 & 3.644  \\ 
             	 & 0.144  & 0.331  & 0.003  & 0.004  & 0.204  & 0.182  & 0.243  & 0.176  & 0.105  & 0.100 & 0.130  \\  
NGC~5046       	 & 1.539  & 5.406  & 0.155  & 0.320  & 4.779  & 2.843  & 2.533  & 1.855  & 0.876  & 0.839 & 3.584    \\ 
               	 & 0.260  & 0.552  & 0.006  & 0.007  & 0.254  & 0.278  & 0.313  & 0.233  & 0.179  & 0.168 & 0.169    \\
NGC~5049       	 & 1.443  & 5.274  & 0.140  & 0.302  & 4.507  & 2.873  & 2.524  & 1.655  & 0.829  & 0.812 & 3.487    \\ 
               	 & 0.168  & 0.355  & 0.004  & 0.004  & 0.164  & 0.178  & 0.201  & 0.149  & 0.116  & 0.108 & 0.107    \\
\hline
{\bf Field}&&&&&&&&&&&\\
NGC~2502       	 & 1.230  &        & 0.155  & 0.319  & 4.991  & 2.861  & 2.511  & 1.661  & 0.865  & 0.837 & 3.661  \\ 
             	 & 0.166  &        & 0.003  & 0.004  & 0.155  & 0.169  & 0.190  & 0.140  & 0.106  & 0.099 & 0.104  \\  
NGC~3203       	 & 2.029  & 6.158  & 0.130  & 0.284  & 4.144  & 3.061  & 2.791  & 1.850  & 0.995  & 0.990 & 3.482  \\ 
             	 & 0.143  & 0.306  & 0.003  & 0.004  & 0.145  & 0.157  & 0.177  & 0.133  & 0.103  & 0.096 & 0.093  \\  
NGC~6684        	 & 2.104  & 6.081  & 0.122  & 0.273  & 3.957  & 2.924  & 2.716  & 1.890  & 1.083  & 1.029 & 3.340    \\
               	 & 0.110  & 0.233  & 0.002  & 0.003  & 0.108  & 0.118  & 0.131  & 0.097  & 0.075  & 0.070 & 0.069    \\
\hline
{\bf Compact Group}&&&&&&&&&&&\\
HCG4C	         & 2.279  & 6.266  & 0.117  & 0.262  & 4.162  & 3.125  & 2.415  & 1.715  & 1.022  & 0.978 & 3.395  \\  
             	 & 0.186  & 0.412  & 0.004  & 0.005  & 0.193  & 0.211  & 0.241  & 0.177  & 0.128  & 0.122 & 0.120  \\
HCG16X           & 5.448  & 2.065  & 0.026  & 0.091  & 2.175  & 1.046  & 0.700  & 0.503  & 0.635  & 0.566 & 1.378  \\
                 & 0.201  & 0.493  & 0.005  & 0.006  & 0.227  & 0.261  & 0.300  & 0.223  & 0.170  & 0.163 & 0.173  \\
HCG22A           & 1.431  &        & 0.170  & 0.333  & 4.935  & 2.955  & 2.475  & 1.795  & 0.992  & 0.895 & 3.660  \\
                 & 0.087  &        & 0.002  & 0.002  & 0.108  & 0.112  & 0.137  & 0.097  & 0.059  & 0.055 & 0.072  \\
HCG22D           & 1.557  & 5.145  & 0.160  & 0.311  & 4.248  & 2.751  & 2.408  & 1.833  & 1.197  & 0.947 & 3.310  \\
                 & 0.187  & 0.399  & 0.004  & 0.004  & 0.189  & 0.197  & 0.229  & 0.168  & 0.117  & 0.112 & 0.122  \\
HCG22E           & 1.480  &        & 0.111  & 0.265  & 4.082  & 2.680  & 2.059  & 1.736  & 1.365  & 0.769 & 3.110  \\
                 & 0.187  &        & 0.004  & 0.005  & 0.183  & 0.199  & 0.226  & 0.165  & 0.120  & 0.114 & 0.121  \\
HCG22X           & 1.593  & 5.666  & 0.143  & 0.308  & 4.927  & 2.812  & 2.528  & 2.041  & 1.015  & 0.931 & 3.627  \\
                 & 0.202  & 0.460  & 0.004  & 0.005  & 0.235  & 0.240  & 0.294  & 0.224  & 0.151  & 0.146 & 0.155  \\
HCG25F           & 2.235  & 4.904  & 0.073  & 0.216  & 3.292  & 2.911  & 2.342  & 1.651  & 0.943  & 0.541 & 2.940  \\
                 & 0.223  & 0.492  & 0.005  & 0.006  & 0.238  & 0.258  & 0.296  & 0.221  & 0.174  & 0.170 & 0.153  \\
HCG32A           & 1.701  & 5.589  & 0.140  & 0.299  & 4.777  & 2.693  & 2.100  & 1.722  & 0.985  & 0.812 & 3.384 \\
                 & 0.168  & 0.370  & 0.004  & 0.004  & 0.176  & 0.186  & 0.216  & 0.201  & 0.118  & 0.112 & 0.159 \\
HCG32B           & 1.813  & 5.211  & 0.121  & 0.270  & 4.666  & 2.695  & 2.248  & 1.757  & 1.120  & 0.754 & 3.396  \\
                 & 0.218  & 0.477  & 0.005  & 0.005  & 0.223  & 0.241  & 0.278  & 0.205  & 0.154  & 0.148 & 0.150  \\
HCG32D           & 1.648  & 5.622  & 0.130  & 0.276  & 4.638  & 3.107  & 2.754  & 1.913  & 1.122  & 0.953 & 3.687  \\
                 & 0.215  & 0.463  & 0.004  & 0.005  & 0.216  & 0.233  & 0.266  & 0.197  & 0.146  & 0.139 & 0.140  \\
HCG40A           &        &        & 0.146  & 0.306  & 4.594  & 2.390  & 2.492  & 1.909  & 0.915  & 0.819 & 3.349  \\
                 &        &        & 0.002  & 0.002  & 0.096  & 0.103  & 0.118  & 0.087  & 0.064  & 0.062 & 0.064  \\
HCG40B       	 & 1.919  & 5.346  & 0.117  & 0.272  & 4.320  & 2.830  & 2.641  & 1.689  & 0.928  & 0.829 & 3.438  \\  
             	 & 0.172  & 0.381  & 0.004  & 0.004  & 0.180  & 0.201  & 0.235  & 0.171  & 0.125  & 0.120 & 0.121  \\  
HCG42A           & 1.337  & 5.321  & 0.160  & 0.314  & 4.523  & 2.827  & 2.367  & 1.570  & 0.852  & 0.814 & 3.427    \\
               	 & 0.227  & 0.549  & 0.006  & 0.007  & 0.293  & 0.325  & 0.389  & 0.300  & 0.260  & 0.254 & 0.201    \\
HCG42C           & 1.832  & 5.105  & 0.125  & 0.275  & 3.927  & 2.079  & 2.285  & 1.694  & 0.602  & 0.753 & 2.927    \\
               	 & 0.263  & 0.624  & 0.007  & 0.008  & 0.328  & 0.377  & 0.435  & 0.333  & 0.303  & 0.294 & 0.229    \\
HCG62:ZM19       & 1.616  & 4.224  & 0.119  & 0.280  & 4.563  & 2.487  & 1.766  & 1.629  & 0.818  & 0.423 & 3.115    \\
               	 & 0.225  & 0.495  & 0.005  & 0.006  & 0.234  & 0.254  & 0.292  & 0.217  & 0.160  & 0.154 & 0.163    \\
HCG86A         	 &        & 5.087  & 0.156  & 0.320  & 5.225  & 2.655  & 2.661  & 1.656  & 0.802  &       & 3.727  \\
               	 &        & 0.426  & 0.004  & 0.005  & 0.270  & 0.253  & 0.346  & 0.256  & 0.188  &       & 0.178  \\
HCG86B    	 &        & 4.799  & 0.118  & 0.269  & 4.364  & 2.645  & 2.363  & 1.505  & 0.591  &       & 3.306  \\       
               	 &        & 0.406  & 0.004  & 0.005  & 0.239  & 0.241  & 0.318  & 0.233  & 0.176  &       & 0.160  \\
\hline
\end{tabular}
\end{center}
\caption{Lick indices (as Table \ref{indices1}).}\label{indices2} 
\normalsize
\end{table*}

\begin{table*}
\footnotesize
\begin{center}
\begin{tabular}{lccccccccccc}
\hline
Galaxy & H$\delta_A$ & H$\delta_F$ & CN$_1$ & CN$_2$ & Ca4227 &
H$\gamma_A$ & H$\gamma_F$ & Fe4383 & Ca4455 & Fe4531 & C$_2$4668 \\
\hline
{\bf Spiral Bulges}&&&&&&&&&&&\\
HCG4A            &       &        &-0.038  &-0.028  & 0.319  &        &        & 2.343  & 0.612  & 1.427 & 1.570  \\
                 &       &        & 0.006  & 0.007  & 0.107  &        &        & 0.268  & 0.134  & 0.208 & 0.314  \\
HCG14A      	 &       &        &        &        & 0.964  &        &        & 3.772  & 1.058  & 2.719  & 5.076   \\ 
             	 &       &        &        &        & 0.177  &        &        & 0.444  & 0.228  & 0.334  & 0.498   \\ 
HCG14B   	 &       & -0.323 & 0.056  & 0.082  & 1.113  &        & -1.802 & 4.541  & 1.387  & 3.396  & 5.191   \\ 
             	 &       & 0.240  & 0.009  & 0.010  & 0.153  &        & 0.193  & 0.369  & 0.184  & 0.252  & 0.403   \\ 
HCG16A           &       & 1.980  & -0.008 & 0.031  & 0.665  & -1.870 & 0.742  & 3.997  & 1.271  & 2.298  & 5.411   \\ 
             	 &       & 0.111  & 0.004  & 0.005  & 0.080  & 0.144  & 0.088  & 0.192  & 0.098  & 0.144  & 0.213   \\ 
HCG16B        	 &-2.807 &        & 0.088  & 0.115  & 1.046  &        &        & 5.519  & 1.558  & 3.313  & 6.742   \\ 
             	 &0.293  &        & 0.007  & 0.008  & 0.119  &        &        & 0.270  & 0.139  & 0.199  & 0.286   \\ 
HCG22B           &2.700  & 2.239  & -0.083 & -0.038 & 1.207  & -1.383 & 1.471  & 3.052  & 1.032  & 2.849  & 4.179   \\ 
                 &0.426  & 0.286  & 0.012  & 0.014  & 0.202  & 0.410  & 0.231  & 0.535  & 0.273  & 0.406  & 0.623   \\ 
HCG25B           &       &        & 0.016  & 0.048  & 1.066  &-4.825  &-1.129  & 4.338  & 0.981  & 2.603 & 4.622   \\ 
                 &       &        & 0.012  & 0.014  & 0.209  & 0.399  & 0.252  & 0.489  & 0.246  & 0.351 & 0.504   \\ 
HCG40D           &1.574  &        & -0.034 & -0.015 & 0.577  &        &        & 3.382  & 0.905  & 2.007  & 4.017  \\
                 &0.303  &        & 0.008  & 0.010  & 0.150  &        &        & 0.367  & 0.188  & 0.284  & 0.412  \\
HCG62:ZM22 	 &-1.335 & 0.412  & -0.011 & 0.013  & 0.887  &        &        & 4.778  & 1.607  & 2.928  & 5.061   \\ 
             	 &0.547  & 0.359  & 0.013  & 0.016  & 0.235  &        &        & 0.553  & 0.246  & 0.407  & 0.595   \\ 
\hline
\end{tabular}
\end{center}
\caption{Lick indices (as Table \ref{indices1}).}\label{indices3} 
\normalsize
\end{table*}

\begin{table*}
\footnotesize
\begin{center}
\begin{tabular}{lccccccccccc}
\hline
Galaxy &H$\beta$ & Fe5015 & Mg$_1$ & Mg$_2$ &  Mgb & Fe5270 & Fe5335 &
Fe5406 & Fe5709 & Fe5782 & [MgFe] \\
\hline
{\bf Spiral Bulges}&&&&&&&&&&&\\  
HCG4A            &        &        & 0.084  & 0.155  & 2.758  & 1.498  & 1.294  & 0.969  & 0.404  & 0.758 & 1.962  \\
                 &        &        & 0.003  & 0.003  & 0.122  & 0.139  & 0.157  & 0.116  & 0.090  & 0.085 & 0.092  \\
HCG14B   	 & 1.507  &        & 0.122  & 0.271  & 4.185  & 2.370  & 2.090  & 1.386  & 0.774  & 1.059 & 3.055  \\      
             	 & 0.160  &        & 0.004  & 0.004  & 0.162  & 0.180  & 0.202  & 0.151  & 0.117  & 0.109 & 0.110  \\  
HCG14A      	 &        &        & 0.086  & 0.207  & 3.444  & 2.258  & 2.008  & 1.453  & 0.732  & 0.946 & 2.710  \\   
             	 &        &        & 0.004  & 0.005  & 0.202  & 0.221  & 0.250  & 0.184  & 0.142  & 0.133 & 0.132  \\  
HCG16A           &        &        & 0.097  & 0.217  & 3.374  & 2.430  & 2.151  & 1.406  & 0.899  & 0.941 & 2.780  \\
             	 &        &        & 0.002  & 0.002  & 0.084  & 0.092  & 0.106  & 0.080  & 0.060  & 0.055 & 0.055  \\  
HCG16B        	 &        &        & 0.149  & 0.281  &        & 2.773  & 2.486  & 1.765  & 1.183  & 0.970 &    \\ 
             	 &        &        & 0.002  & 0.003  &        & 0.118  & 0.132  & 0.098  & 0.073  & 0.069 &    \\  
HCG22B           & 3.105  & 4.612  & 0.060  & 0.174  & 2.708  & 2.179  & 1.759  & 1.175  & 0.983  & 0.743 & 2.309  \\
                 & 0.244  & 0.553  & 0.006  & 0.007  & 0.269  & 0.299  & 0.341  & 0.255  & 0.199  & 0.189 & 0.176  \\
HCG25B           &        &        & 0.106  & 0.240  & 4.296  & 2.310  & 2.020  & 1.383  & 0.863  & 0.616 & 3.050  \\
                 &        &        & 0.004  & 0.005  & 0.175  & 0.194  & 0.217  & 0.159  & 0.115  & 0.109 & 0.124  \\
HCG40D           &        &        & 0.077  & 0.185  & 3.419  & 1.915  & 1.632  & 1.054  & 0.641  & 0.619 & 2.462  \\
                 &        &        & 0.003  & 0.004  & 0.161  & 0.181  & 0.203  & 0.150  & 0.113  & 0.108 & 0.117  \\
HCG62:ZM22 	 & 2.213  & 4.810  & 0.084  & 0.227  & 3.475  & 2.625  & 2.321  & 1.596  & 0.967  & 0.941 & 2.931  \\  
             	 & 0.229  & 0.497  & 0.005  & 0.006  & 0.234  & 0.254  & 0.287  & 0.213  & 0.165  & 0.156 & 0.150  \\  
\hline
\end{tabular}
\end{center}
\caption{Lick indices (as Table \ref{indices1}).}\label{indices4}               
\normalsize
\end{table*}

\section{Aperture and abundance ratio correction}
\label{LARC}
This section details the correction of derived parameters for the trends
seen in the solar neighbourhood abundance ratios. 
Data from the studies of PS02 and T00 are included for comparison purposes. 
Interpretation of the data is discussed in the main body of the paper (Section
\ref{agez}).

\subsection{Local abundance ratio correction}
Estimates of [E/Fe] in the three studies are derived by comparison of Lick
indices in galaxies to SSP estimates based on collections of local stars. 
These estimates are clearly measurements of the difference between 
the [E/Fe] in the observed galaxies and that in local stars. This is 
important, as the abundance ratios of key elements (e.g. C, O and Mg)
in stars in the solar neighbourhood have been shown to vary with
[Fe/H], (Nissen \& Edvardsson 1992; Edvardsson \etal 1993; Felzing \& Gustafsson 1998; 
Gustafsson \etal 1999; Bensby \etal 2003). This effect must 
clearly be accounted for if we are to fully calibrate our estimates of [E/Fe] to 
the solar value. Indeed, it will be seen that, rather than being just a
minor correction, consideration of this effect has a significant
impact on correlations within the data. Consequently, this is a most
important (and sometimes overlooked) aspect of the full calibration to the
Lick system.

The most important elements for the analysis presented here are C, O, Mg,
as these include the main elements upon which the enhancement 
sensitive indices used in this work (CN$_1$, CN$_2$, Ca4227, 
C$_2$4668, Mg$_1$, Mg$_2$ and Mgb) are dependent, and Fe to which
the remainder of metallicity sensitive indices are most sensitive. 
We shall therefore concentrate on these elements in the discussion that follows.

Early studies of the pattern of abundance ratios in local stars (e.g. Nissen 
\& Edvardsson 1992; Edvardsson \etal 1993) found that, in stars with [Fe/H]$<$0, 
strong anti-correlations exist between the abundances of various elements and [Fe/H]. 
The E group elements C, O 
and Mg were all shown to exhibit such a trend with a gradient $\sim$ --0.5 
dex/dex. However, these studies suggested that, at higher [Fe/H] these anti-correlations 
disappeared, with all stars with [Fe/H]$>$0 having solar abundance ratios.
More recent work (incorporating non-LTE considerations) has changed this 
picture somewhat. The studies of Felzing \& Gustafsson (1998), Gustaffsson 
\etal (1999) and Bensby \etal (2003) showing 
that the trend seen at low metallicities continues at [Fe/H]$>$0, at least for 
O and C. It should be noted 
that C is the most important element in the estimation of indices in enhanced 
SSPs because, of the seven enhancement sensitive indices listed above, 
five are more sensitive to C than any other element. On the other hand, O is 
the element that dominates the E group in terms of mass. These two elements
therefore dominate the mass of the E group elements as well as the sensitivities 
of the indices form which [E/Fe] is derived.

\begin{figure*}
\centerline{\psfig{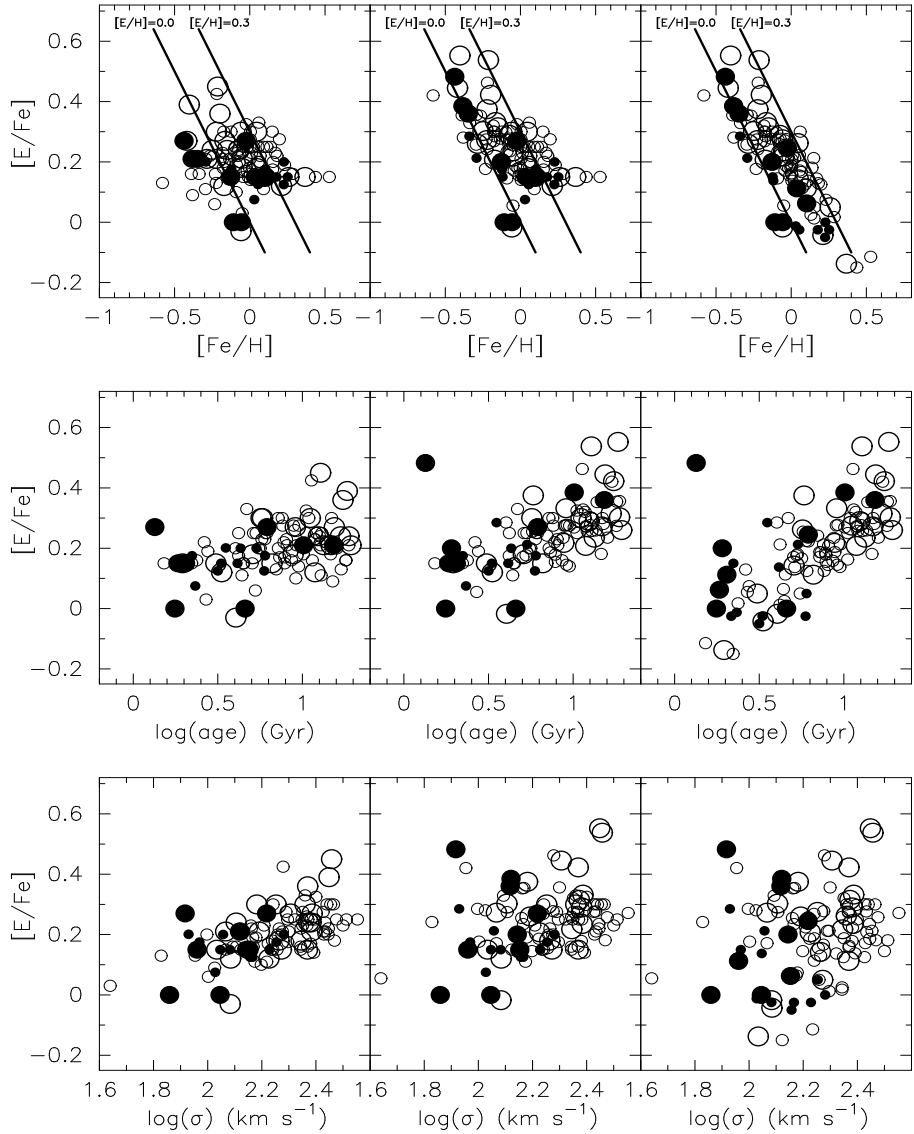}}
\caption{\label{larc1} Derived parameters for the combined sample are 
plotted before and after correction for the local abundance ratio pattern. 
Spiral bulges are shown as filled circles. Large symbols are the present
work while small symbols are the data from T00 and PS02. The left-hand panels
show `raw' data. Central panels show data after galaxies with [Fe/H]$<$0
have been corrected for the local abundance ratio trends. Right-hand panels 
show data when all galaxies have been corrected. Lines of constant [E/Fe] are 
shown in [E/Fe]--[Fe/H] plots.}
\end{figure*}

\begin{figure*}
\centerline{\psfig{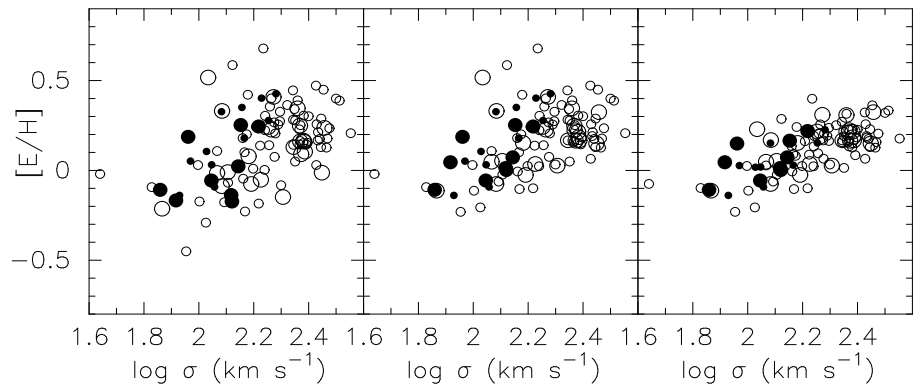}}
\caption{\label{larc2} Symbols as Fig. \ref{larc1}. The correlations between
[E/H] and log $\sigma$ are the same in early- and late-type galaxies}
\end{figure*}

In the light of these findings the correction for the trend of
decreasing [E/Fe] with [Fe/H], which was not included for [Fe/H]$>$0 
in many preceding studies (including PS02), should be extended into this 
regime. To illustrate the effects of extending this correction Figures 
\ref{larc1} and \ref{larc2} show the results of terminating at [Fe/H]=0
and also continuing to higher [Fe/H]. We note that the lowest [Fe/H] in
this study was -0.65 dex, well above the -1 dex
at which the trends of elemental abundance with [Fe/H] change again (Feltzing \&
Gustafsson (1998). It can be seen in the
plots of Figs \ref{larc1} and \ref{larc2} that this correction has profound
effects
on correlations in the data. For instance, weak [E/Fe]--log(age) and
[E/H]--log $\sigma$ correlations are substantially strengthened by the
correction, while an apparent correlation between [E/Fe] and log $\sigma$ is
destroyed. A correlation also appears between [E/Fe]
and [Fe/H] that was not apparent before application of the correction.
This is not as surprising given that the magnitude of the correction is directly
proportional to the [Fe/H] over the range it is applied. However, it is
apparent that the correlation is only strengthened (not created) by the
extension of the corrected regime to high [Fe/H]. It should also be noted
that the extension to high [Fe/H] also has the effect of making the T00 and 
PS02 data consistent with those from the present study.

\begin{table*}
\footnotesize
\begin{center}
\begin{tabular}{lrrrrrr}
\hline
Galaxy        &log(age)$_{raw}$&log(age)$_{corr}$&[Fe/H]$_{raw}$&[Fe/H]$_{corr}$&[E/Fe]$_{raw}$&[E/Fe]$_{corr}$\\  
\hline					                       		                     
{\bf EARLY TYPE} & & & &&&\\		                       		                     
{\bf Loose group} & & & &&&\\		                       		                     
NGC~3305       &   1.225(0.075) &   1.240(0.075) & -0.125(0.062) & -0.198(0.062) & 0.360(0.050) & 0.423(0.050) \\ 
NGC~5046       &   1.250(0.175) &   1.275(0.175) & -0.125(0.113) & -0.249(0.113) & 0.240(0.060) & 0.303(0.060) \\ 
NGC~5049       &   1.150(0.100) &   1.170(0.100) & -0.125(0.100) & -0.224(0.100) & 0.210(0.030) & 0.272(0.030) \\ 
\hline
{\bf Field} & & & &&&\\     		                       		                     
NGC~2502       &   1.250(0.050) &   1.287(0.050) & -0.100(0.062) & -0.286(0.062) & 0.210(0.045) & 0.260(0.045) \\ 
NGC~3203       &   0.500(0.125) &   0.523(0.125) &  0.325(0.125) &  0.211(0.125) & 0.120(0.030) &-0.042(0.030) \\ 
NGC~6684       &   0.250(0.112) &   0.292(0.112) &  0.575(0.150) &  0.367(0.150) & 0.150(0.045) &-0.137(0.045) \\ 
\hline
{\bf Compact group} & & & &&&\\    		                       		                     
HCG4C         &   0.500(0.125) &   0.488(0.125) &  0.200(0.150) &  0.259(0.150) & 0.150(0.045) & 0.050(0.045) \\ 
HCG16X        &  -0.126(0.075) &  -0.111(0.075) & -0.200(0.100) & -0.273(0.100) & 0.060(0.045) & 0.160(0.045) \\
HCG22A        &   1.025(0.050) &   1.047(0.050) &  0.050(0.038) & -0.058(0.038) & 0.300(0.015) & 0.275(0.015) \\
HCG22D        &   1.225(0.100) &   1.225(0.100) & -0.150(0.075) & -0.150(0.075) & 0.240(0.060) & 0.315(0.060) \\
HCG22E        &   1.075(0.150) &   1.074(0.150) & -0.175(0.138) & -0.172(0.138) & 0.120(0.060) & 0.207(0.060) \\
HCG22X        &   1.125(0.100) &   1.125(0.100) & -0.050(0.100) & -0.050(0.100) & 0.210(0.060) & 0.235(0.060) \\
HCG25F        &   0.600(0.188) &   0.607(0.188) & -0.025(0.150) & -0.058(0.150) & 0.030(0.060) &-0.017(0.060) \\
HCG32A        &   1.100(0.150) &   1.095(0.150) & -0.100(0.113) & -0.074(0.113) & 0.240(0.045) & 0.290(0.045) \\ 
HCG32B        &   0.950(0.175) &   0.945(0.175) & -0.075(0.150) & -0.050(0.150) & 0.180(0.045) & 0.218(0.045) \\ 
HCG32D        &   0.825(0.150) &   0.820(0.150) &  0.075(0.113) &  0.099(0.113) & 0.150(0.060) & 0.113(0.060) \\ 
HCG40A        &   0.750(0.088) &   0.756(0.088) &  0.075(0.050) &  0.046(0.050) & 0.300(0.015) & 0.263(0.015) \\ 
HCG40B        &   0.700(0.175) &   0.706(0.175) &  0.075(0.125) &  0.047(0.125) & 0.180(0.045) & 0.143(0.045) \\ 
HCG42A        &   1.250(0.075) &   1.266(0.075) & -0.325(0.062) & -0.403(0.062) & 0.390(0.060) & 0.553(0.060) \\ 
HCG42C        &   0.750(0.300) &   0.764(0.300) & -0.150(0.175) & -0.222(0.175) & 0.300(0.100) & 0.375(0.100) \\ 
HCG62:ZM19    &   1.175(0.100) &   1.189(0.100) & -0.350(0.088) & -0.419(0.088) & 0.270(0.055) & 0.445(0.055) \\
HCG86A        &   1.100(0.100) &   1.108(0.100) & -0.175(0.087) & -0.213(0.087) & 0.450(0.050) & 0.537(0.050) \\ 
HCG86B        &   0.950(0.113) &   0.958(0.113) & -0.125(0.100) & -0.165(0.100) & 0.270(0.045) & 0.333(0.045) \\ 
\hline			                        		                       		                     
{\bf SPIRAL BULGES}  & & & &&&\\		                       	     	      	     	        	     	            	            
HCG4A         &   0.125(0.087) &   0.127(0.087) & -0.425(0.138) & -0.437(0.138) & 0.270(0.030) & 0.483(0.030) \\
HCG14A        &   0.301(0.250) &   0.309(0.250) &  0.075(0.200) &  0.037(0.200) & 0.150(0.075) & 0.113(0.075) \\
HCG14B        &   1.175(0.100) &   1.184(0.100) & -0.300(0.075) & -0.347(0.075) & 0.210(0.045) & 0.360(0.045) \\ 
HCG16A        &   0.250(0.150) &   0.264(0.150) &  0.175(0.087) &  0.103(0.087) & 0.150(0.030) & 0.063(0.030) \\
HCG16B        &   0.775(0.038) &   0.790(0.038) &  0.050(0.050) & -0.026(0.050) & 0.270(0.015) & 0.245(0.015) \\
HCG22B        &   0.225(0.051) &   0.247(0.051) &  0.000(0.113) & -0.108(0.113) & 0.000(0.060) & 0.000(0.060) \\
HCG25B        &   1.000(0.175) &   1.007(0.175) & -0.350(0.137) & -0.383(0.137) & 0.210(0.045) & 0.385(0.045) \\
HCG40D        &   0.276(0.075) &   0.281(0.075) & -0.100(0.125) & -0.127(0.125) & 0.150(0.045) & 0.200(0.045) \\ 
HCG62:ZM22    &   0.650(0.250) &   0.661(0.250) &  0.000(0.225) & -0.056(0.225) & 0.000(0.060) & 0.000(0.060) \\ 
\hline									   
\end{tabular}								   
\end{center}								   
\caption{Results of age/metallicity determinations. Both raw data and data 
corrected for aperture effects and local abundance ratio patterns (Appendix
\ref{LARC}) are given.}
\label{dervals}			   
\normalsize								   
\end{table*}

\end{appendix}
\label{lastpage}
\end{document}